\documentclass[onecolumn,groupedaddress,nofootinbib,aps,prb]{revtex4-1}
\usepackage{amsmath}
\usepackage{epsfig,color}
\usepackage{graphicx}
\usepackage{dcolumn}
\usepackage{bm}
\usepackage{multirow}
\usepackage{bm}
\usepackage{enumerate}
\usepackage{graphicx}
\usepackage{verbatim}
\usepackage{hyperref}
\usepackage[utf8]{inputenc}
\usepackage[english]{babel}
\usepackage{amssymb,color}
\usepackage{amsfonts}
\usepackage{graphicx}
\usepackage{dcolumn}
\usepackage{float}
\usepackage{bm}
\def\beq{\begin{equation}}
\def\eeq{\end{equation}}
\def\bea{\begin{eqnarray}}
\def\eea{\end{eqnarray}}

\begin{document}
\title{Dynamic properties of superconductors:  Anderson-Bogoliubov mode and Berry phase in BCS and BEC regimes}

\author{Dmitry Mozyrsky} \email{mozyrsky@lanl.gov}
\address{Theoretical Division (T-4), Los Alamos National Laboratory, Los Alamos NM 87545}

\author{Andrey V. Chubukov} \email{achubuko@umn.edu}
\address{Department of Physics, University of Minnesota, Minneapolis MN 55455}

\date{\today}
\begin{abstract}
We analyze the evolution of the  dynamics of a neutral s-wave superconductor between BCS and BEC regimes. We  consider 2d case, when BCS-BEC crossover occurs already at weak coupling as a function of the ratio of the two scales -- the Fermi energy $E_F$ and the bound state energy for two fermions in a vacuum, $E_0$.  BCS and BEC limits correspond to  $E_F \gg E_0$ and $E_F \ll E_0$, respectively. The chemical potential $\mu = E_F-E_0$ changes the sign between the two regimes. We use the effective action approach, derive the leading terms in the expansion of the effective action in the spatial and time derivative of the slowly varying superconducting order parameter $\Delta (r, \tau)$, and express the action in terms of derivative of the phase $\phi (r,\tau)$ of $\Delta (r, \tau) = \Delta e^{i\phi (r, \tau)}$.  The action contains $(\nabla \phi)^2$ and ${\dot \phi}^2$ terms, which determine the dispersion of collective phase fluctuations, and $i \pi  A {\dot \phi}$ term. For continuous $\phi (r,\tau)$, the latter  reduces to the contribution from the boundary and does not affect the dynamics. We show that this longwavelength action does not change through BCS-BEC crossover.
We apply our approach to a moving vortex,  for which $\phi$ is singular at the center of the vortex core, and $i \pi A_{vort} {\dot \phi}$ term affects  vortex dynamics.  We find that this term has  two contributions. One comes from the states away from the vortex core and has $A_{vort,1} = n/2$, where  $n$ is the fermion density. The other comes from  electronic states inside the vortex core and has $A_{vort,2} = -n_0/2$, where $n_0$ is the fermion density at the vortex core. This last term comes  from the continuous part of the electronic spectrum and has no contribution from discrete levels inside the core; it also does not change if we add impurities.   We interpret this term as the contribution to vortex dynamics in the continuum limit, when the spacing between energy levels $\omega$ is set to zero, while fermionic lifetime $\tau$ can be arbitrary.  The total $A_{vort} = (n-n_0)/2$ determines the transversal force acting on the vortex core, $\pi A_{vort} {\dot {\bf R}}\times \hat z$, where ${\dot {\bf R}}$ is the velocity of the vortex core and $\hat z$ a unit vector perpendicular to the 2d sample. The difference $(n-n_0)/2$ changes through the BEC-BCS crossover as $n_0$ nearly compensates $n$ in the BCS regime, but vanishes in the BEC regime.
\end{abstract}
\maketitle

\section{Introduction}

The evolution of the static properties of a superconductor between BCS regime, when bound pairs of fermions condense immediately once they form, and Bose-Einstein condensation (BEC)  regime, when bound pairs of fermions form at a higher $T_{ins}$ and condense at a smaller $T_c$, has been extensively discussed in the condensed matter context~\cite{Miyake1983,Nozieres85,Mohit89,zwerger,Ohashi,DeMelo1993,ioffe97,tobij2000,lara2001,DeMelo2006,lara2009,Innocenti2010,ranninger, levchenko2011,lara2012,Guidini2014,Mohit2015,kagan,Ohashi,Guidini2014,Aleiner15} and  also for optical lattices of ultracold atoms \cite{s_1,jin}. Experimental evidence for preformed pairs has been  reported for high-$T_c$ cuprates~\cite{cuprates_pp} and, more recently, for Fe-based superconductor FeSe$_{1-x}$Te$_x$ (Ref. \cite{shibauchi}).

In this paper, we discuss the evolution of the dynamic  properties of a neutral superconductor between BCS and BEC regimes. We consider  a 2d superconductor and for definiteness focus on s-wave gap symmetry and assume  Galilean invariance,  i.e., assume rotational symmetry and $k^2/(2m)$ fermionic dispersion.  Extensions to non-s-wave pairing and lattice systems are straightforward.  We consider 2d case because in 2d, BCS-BEC crossover can be analyzed already within weak coupling, when calculations are under control.  Indeed, in 2d systems, two fermions form a bound state already at arbitrary small attraction $g$. (In  3d systems, the bound state of two fermions in a vacuum emerges only once the interaction exceeds a certain cutoff, generally of the order of fermionic bandwidth~\cite{DeMelo1993}.) Such a bound state has energy $2E_0=2\Lambda e^{-2/(N_0g)}$, where $N_0=m/(2\pi)$ is the free particle density of states per spin in 2d and $\Lambda$ is the upper cutoff for the attraction \cite{Miyake1983,Mohit89,chubukov_17}.

The crossover between BCS and BEC regimes occurs as a function of $E_0/E_F$.  For $E_F >> E_0$ the system is in BCS regime, and bound pairs condense almost instantly after they form at $T_{ins}  \sim (E_F E_0)^{1/2}$.    For $E_0 >> E_F$, bounds pairs form at $T_{ins} \sim E_0/\log{E_0/E_F}$ and condense at a much smaller $T_c \sim E_F$, leaving a wide intermediate region of preformed pair behavior.   The chemical potential $\mu$ at $T=0$ changes sign between the two regimes: $\mu = E_F - E_0$.

We use the effective action approach, and expand the action in terms of time derivatives of the slowly varying order parameter $\Delta (r, \tau)$.  We obtain the generic expressions for the terms up to second order in spatial and time derivatives, in terms of the eigenfunctions of the Bogolubov-De Gennes equation, and then apply the results to the case when the variations of the order parameter predominantly occur via variations of its phase, i.e., $\Delta (r, \tau) \approx \Delta e^{i \phi (r, \tau)}$.  We obtain the action in terms of spatial and time derivatives of $\phi$.  The action contains the conventional terms $\nabla^2 \phi$ and $\partial^2 \phi/\partial \tau^2$, which fully describe the dynamics when $\phi$ is a continuous function of a coordinate and time
 \begin{equation}
\label{wed_3_1}
{\cal S}_{cont} \propto  \sum_{q,\Omega} |\phi_{q,\Omega}|^2 \left(\Omega^2 - q^2 \frac{v^2_F}{2} \right).
\end{equation}
The form of ${\cal S}_{cont}$ does not change between BCS and BEC limits, and the velocity of phase fluctuations  remains $v_F/\sqrt{2}$ through the crossover.

Using our approach we also study the dynamics of vortices and, in particular, the transverse force acting on a slowly moving vortex. Such force is typically attributed to the terms in the effective action that are linear in time derivatives of the phase, i.e., proportional to $\int dr d \tau  {\dot \phi}$. This term is often referred to in the literature as Berry phase term~\cite{thouless,otterlo,jacek,simanek_2,stone_gaitan,stone,volovik_book,we_1}. It reduces to the contribution from a boundary and does not contribute to the dynamics if $\phi$ is well defined at any $r$ and $\tau$. However, for vortices, as well as for other topological defects, such as phase slips ~\cite{phase_slips}, such term does contribute to the dynamics due to non-analytic behaviour of $\phi$ at the center of the vortex core, and gives rise to an effective transverse force acting on a vortex~\cite{bardeen_2,kopnin,kopninvolovik,volovik,volovik_book,stone_gaitan,thouless,jacek,gaitan,simanek_2}.
The action associated with this transverse force can be written as
\bea
S^{vort}_{Berry} = i \pi \pi A_{vort} \int d t \left( X(t) {\dot Y} (t) -  Y(t) {\dot X} (t) \right)\,, \nonumber \\
\label{lll}
\eea
with $X$ and $Y$ being the coordinates of the vortex core.

We show that the prefactor $A_{vort}$ has two contributions, $A_{vort} = A_{vort,1} + A_{vort,2}$.
The first one is the hydrodynamic contribution, associated with superfluid motion of fermions at the peripheral region of the vortex (this term is often termed as Magnus force).  We find $A_{vort,1} = n/2$, where $n = 2N_0 E_F$ is the actual fermionic density. Another contribution  is a reaction force from normal fermions at the vortex core. For this term we find $A_{vort,2} = -n_0/2$, where   $n_0 = 2 N_0 \mu$ is the density of free fermions with the same chemical potential $\mu = E_F - E_0$ (same as the density of fermions inside the vortex core). We find, within our approach,  that it comes from the term in the action at $\Delta \to 0$ and does not  change if we add impurities and make fermionic lifetime $\tau$ finite.  At $\Delta \to 0$,  the distance between energy levels in the core $\omega \sim \Delta^2/E_F$ (Refs. \cite{caroli,bardeen, kopninvolovik} ) vanishes and the fermionic spectrum in the vortex core becomes continuous. The Berry term (\ref{lll}) with $A_{vort} = (n-n_0)/2$  then should be viewed as the part of the action for the vortex motion in the continuous limit. In the notations of Refs. \cite{kopninvolovik, volovik} this corresponds to $\omega \tau \ll 1$.

The interplay between the Magnus and the reaction forces is different in BCS and BEC limits, i.e., for $E_F > E_0 $ and for $E_F < E_0$.
In the BCS regime  $E_F > E_0$ the difference  $n-n_0 = 2N_0 E_0 \ll n$,  i.e., these two forces nearly cancel each other. The resulting  $A_{vort} = N_0 E_0$.  In the BEC regime, $\mu <0$, i.e., all states of free fermions are above the chemical potential and therefore are empty.  Then $n_0=0$, and only Magnus force contributes to $A_{vort} =  A_{vort,1} = n/2 = N_0 E_F$. The vanishing of $n_0$ once $\mu$ becomes negative is consistent with the generic reasoning in Ref.~\cite{volovik} that free-fermion contribution to $A_{vort}$ vanishes once the system undergoes a (fictitious) Lifshitz transition, in which the (fictitious) Fermi surface of free fermions with renormalized $\mu$ disappears. In our case, this happens once $E_0$ becomes larger than $E_F$.

Our microscopic results agree with earlier works~\cite{volovik,volovik_1,otterlo_2,kopninvolovik}, which argued on general grounds  that at $\omega \tau \ll 1$, $A_{vort}$ should be equal to $(n-n_0)/2$. These  researchers also speculated that there should be a contribution to $A_{vort}$ from discrete levels in the vortex core, for which level spacing $\omega$ s finite,  and that in the limit $\omega \tau \gg 1$, the total contribution from the vortex core $A_{vort,2}$ should vanish, i.e., the total $A_{vort} = A_{vort,1} = n/2$ (Refs.  \cite{kopninvolovik,stone,volovik_book,volovik,otterlo,otterlo_2}). We  didn't find in our microscopic approach the contribution to $A_{vort}$ from  discrete levels in the vortex core in the terms in the action, in which  $\Delta$ is finite. It remains to be seen whether  such contribution emerges if one goes beyond the approximations we made in the derivation of the action.

The paper is organized as follows: In the  next section we introduce the effective action of a  superconductor in terms of its fluctuating order parameter $\Delta({\bf r},\tau)$. In Section 3  we develop a systematic expansion of the action  in terms of (imaginary) time derivatives  of the order parameter. We evaluate the zeroth order term  and obtain the condensation energy of a superconductor. We then obtain terms with one and two time derivatives, and express them in terms of eigenfunctions of the Bogoliubov-de Gennes equation.  We next focus on the small wavelength limit and express the action in terms of spatial and time derivatives of the phase of a superconducting order parameter, including the term, linear in time derivative. This last  term becomes meaningful Berry phase term when the phase of the superconducting order parameter is not defined globally, which is the case of a vortex.  In Section 4 we compute the effective action for a moving vortex in a neutral s-wave superconductor in 2d. Section 5 is the summary of our results.

\section{General formulation}

The effective action for an order parameter of an s-wave superconductor can be obtained by departing from a microscopic model with local four-fermion attractive interaction $-g$ ($g>0$) and introducing the pairing field $\Delta (r, \tau)$ to decoupling four-fermion interaction via Hubbard-Stratonovich transformation~\cite{Hubb_Str}. This procedure is well documented (see e.g. Ref. \cite{pra}), and we just quote the results.

The partition function $Z$ is expressed via the integral over the Grassmann fields as
\beq
Z = \int d \psi d {\bar \psi} e^{-S[\psi, {\bar \psi}]}
\eeq
where  $\psi = \psi_\alpha ({\bf r}, \tau)$ and ${\bar \psi} = {\bar \psi}_{\alpha} ({\bf r},\tau)$  are spin-full coordinate and time dependent Grassmann fields, and
\beq
S [\psi, {\bar \psi}] = \int d{\bf r} d \tau \left({\bar \psi}_\alpha ({\bf r}, \tau) \partial_\tau \psi_\alpha ({\bf r}, \tau)+H[\psi, {\bar \psi}]\right)
\eeq
Here $\tau$ is the  imaginary (Matsubara) time $\tau = i t$ and
\beq
H[\psi, {\bar \psi}] = \left[{\bar \psi}_\sigma({\bf r},\tau)\left(-{\nabla^2\over 2m}-\mu\right)\psi_\sigma ({\bf r},\tau)\right]  - g {\bar \psi}_{\uparrow} ({\bf r},\tau) {\bar \psi}_{\downarrow} ({\bf r},\tau)
\psi_{\downarrow} ({\bf r},\tau) \psi_{\uparrow} ({\bf r},\tau)
\eeq
The four-fermion interaction is decoupled by Hubbard-Stratonovich transformation
\beq
\mathrm{e}^{\frac{ax^{2}}{2}}=\frac{1}{\sqrt{2\pi a}}\int dy\,\mathrm{e}^{\left(-\frac{y^{2}}{2a}+yx\right)}
\label{Hubbard_Stratononich}
\eeq
In our case we introduce two Hubbard-Stratonovich fields $\Delta ({\bf r}, \tau)$ and $\Delta^* ({\bf r}, \tau)$ and re-write the partition function as
\beq
Z = \int d \psi d {\bar \psi} d \Delta d \Delta^* e^{-S[\psi, {\bar \psi}, \Delta, \Delta^*]},
\eeq
where now
\beq
S [\psi, {\bar \psi},\Delta,\Delta^*] = \int d{\bf r} d \tau \left({\bar \psi}_\alpha ({\bf r}, \tau) \partial_\tau \psi_\alpha ({\bf r}, \tau)+ \frac{|\Delta ({\bf r},\tau)|^2}{g}  +  H[\psi, {\bar \psi},\Delta,\Delta^*]\right)
\eeq
and
\begin{equation}\label{bcs}
H [\psi, {\bar \psi},\Delta,\Delta^*] = \left[{\bar \psi}_\sigma ({\bf r},\tau)\left(-{\nabla^2\over 2m}-\mu\right)\psi_\sigma ({\bf r},\tau) + \Delta({\bf r},\tau){\bar \psi}_{\uparrow} ({\bf r},\tau){\bar \psi}_{\downarrow} ({\bf r},\tau) +\Delta^\ast({\bf r},\tau)\psi_{\downarrow}({\bf r},\tau)\psi_{\uparrow}({\bf r},\tau) \right]\,.
\end{equation}
The action $S [\psi, {\bar \psi},\Delta,\Delta^*]$ can be re-expressed in a more compact form by introducing Gorkov-Nambu spinor
$\psi= \left[\psi_{\uparrow}, {\bar \psi}_{\downarrow}\right]^T$.  Then
\beq
S [\psi, {\bar \psi},\Delta,\Delta^*] = \int d{\bf r} d \tau \frac{|\Delta ({\bf r},\tau)|^2}{g} - \int d{\bf r} d \tau\, {\bar \psi} ({\bf r}, \tau) {\hat G}^{-1} \psi ({\bf r}, \tau),
\eeq
where the ${\hat G}^{-1}$ is an operator
\begin{equation}
{\hat G}^{-1}= -\partial_\tau - {\hat K}({\bf r}) - {\hat \Delta}({\bf r},\tau)\,,
\end{equation}
with
\begin{equation}\nonumber
{\hat K}({\bf r}) =
\begin{bmatrix}
-(1/2m)\nabla^2-\mu & 0 \\
0 & (1/2m)\nabla^2+\mu
\end{bmatrix}\,,
\end{equation}
and
\begin{equation}\nonumber
{\hat \Delta}({\bf r},\tau) =
\begin{bmatrix}
0 & \Delta({\bf r},\tau) \\
\Delta^\ast({\bf r},\tau) & 0
\end{bmatrix}\,.
\end{equation}
The  Green's function for the fermions ${\hat G} ({\bf r}, \tau;{\bf r}',\tau')$ satisfies the following operator identity
\beq
\label{ng}
\big(-\partial_\tau - {\hat K}({\bf r}) - {\hat \Delta}({\bf r},\tau,\lambda)\big){\hat G}({\bf r}, \tau;{\bf r}',\tau')  = \delta({\bf r}-{\bf r}^\prime)\delta(\tau-\tau^\prime)\,.
\end{equation}

Integrating over  $\psi$ and ${\bar \psi}$ we then obtain
\beq
Z = \int d \Delta d \Delta^* e^{-S[\Delta, \Delta^*]}
\eeq
and
\beq
S[\Delta, \Delta^*] = \int d{\bf r} d \tau \frac{|\Delta ({\bf r},\tau)|^2}{g} - Tr \log{{\hat G}^{-1}}
\label{bb1}
\eeq
The logarithm in the effective action can be eliminated by introducing an auxiliary  variable $\lambda$ and  making $\Delta$, and hence $G$, $\lambda$-dependent, subject to $\Delta({\bf r},\tau,1) = \Delta({\bf r},\tau)$ and $\Delta({\bf r},\tau,0)=0$. Indeed,  because ${\hat G}^{-1} (r, \tau) = {\hat G}^{-1}_0 (r, \tau) - {\hat \Delta} (r, \tau)$, we have $\log {\hat G}^{-1} = \log{{\hat G}^{-1}_0}  - \sum_{n=1} ({\hat G}_0 {\hat \Delta})^n/n$. The product  ${\hat G} {\hat \Delta} = \sum_{n=1} ({\hat G}_0 {\hat \Delta})^n$ is represented by the same expansion, but without $1/n$. The $1/n$ can be re-introduced by using the identity
\begin{equation}\nonumber
\int_0^1\,d\lambda\, {\rm Tr}\Big[{\partial{\hat\Delta}(\lambda)\over\partial\lambda}{\hat G}_0\Big({\hat \Delta(\lambda)}{\hat G}_0\Big)^{n-1}\Big]={1\over n}{\rm Tr}\Big[\Big({{\hat\Delta(1)}\hat G}_0\Big)^n\Big].
\end{equation}
Using this trick, we can replace  ${\cal S}$ in (\ref{bb1}) by
\begin{equation}\label{Z}
{\cal S}= \int_{-\infty}^{\infty} d\tau \int d{\bf r} \int_0^1 d\lambda\, {\rm Tr}\,\big[{\hat {\bf G}}_\lambda ({\bf r}, \tau; {\bf r}, \tau){\partial\over\partial\lambda}{\hat \Delta}({{\bf r},\tau,\lambda})\big]
+ \int d \tau \int d{\bf r} \frac{|\Delta ({\bf r},\tau,1)|^2}{g}
+ {\cal S}_{norm},
\end{equation}
where
\beq
{\cal S}_{norm} =- Tr \log{{\hat G}^{-1}_{\Delta \to 0}}.
\label{chn_1}
\eeq
Note that in Eq. (\ref{Z}) the trace is over the Gorkov-Nambu $2\times 2$ matrices only, while in Eq. (\ref{chn_1}) the trace is assumed to be over the infinite-dimensional matrix ${\hat G}^{-1}$ as well as over the Gorkov-Nambu $2\times 2$ structure.

A similar trick  has been used  in Ref.~\cite{agd}, where an  auxiliary variation of the coupling constant was introduced instead of $\lambda-$dependence. Writing the action in the form of Eq. (\ref{Z}) helps with the computations and will also allow us to  establish a connection with the Wess-Zumino formalism, which has been used in earlier works~\cite{volovik,volovik_book,volovik_1} to evaluate the Berry phase and the Magnus force for topological defects, such as vortices.

\section{Adiabatic expansion}

We set ${\hat \Delta}({\bf r},\tau,\lambda)$  to be a slowly varying function of  $\tau$ and expand it around a particular $\tau'$  as
\begin{equation}\label{Delta}
{\hat \Delta}({\bf r},\tau,\lambda) = {\hat \Delta}({\bf r},\tau^\prime,\lambda) + \partial_{\tau^\prime}{\hat \Delta}({\bf r},\tau^\prime,\lambda)(\tau-\tau^\prime) + (1/2)\partial^2
_{\tau^\prime}{\hat \Delta}({\bf r},\tau^\prime,\lambda)(\tau-\tau^\prime)^2 +... \ .
\end{equation}
Consequently, we seek for the solution of Eq. (\ref{ng}) in the form
\begin{equation}\label{G}
{\hat {\bf G}}_\lambda ({\bf r}, \tau; {\bf r}^\prime, \tau^\prime) = {\hat {\bf G}}_\lambda^{(0)} ({\bf r}, \tau-\tau^\prime; {\bf r}^\prime, \tau^\prime) + {\hat {\bf G}}_\lambda^{(1)} ({\bf r}, \tau-\tau^\prime; {\bf r}^\prime, \tau^\prime)+{\hat {\bf G}}_\lambda^{(2)} ({\bf r}, \tau-\tau^\prime; {\bf r}^\prime, \tau^\prime) +... \ ,
\end{equation}
with ${\hat {\bf G}}_\lambda^{(0)}$ being of the order $(\partial_\tau\Delta)^0$, ${\hat {\bf G}}_\lambda^{(1)}$ being of the order $(\partial_\tau\Delta)^1$, etc.
The functional  in Eq. (\ref{Z}) can then be written as a series
\begin{equation}\label{F}
{\cal S}=
{\cal S}_{norm} + {\cal S}_0+{\cal S}_1+{\cal S}_2+ ... = {\cal S}_{norm} + \int d\tau \left[L^{(0)}(\tau) + \int d{\bf r} \frac{|\Delta ({\bf r},\tau,1)|^2}{g}\right] + \int d\tau L^{(1)}(\tau) + \int d\tau L^{(2)}(\tau) + ... \ .
\end{equation}
with
\begin{equation}\label{F0}
L^{(k)}(\tau)=\int d{\bf r} \int_0^1 d\lambda\, Tr\,\big[{\hat {\bf G}}_\lambda^{(k)} ({\bf r}, \tau; {\bf r}, \tau){\partial\over\partial\lambda}{\hat \Delta}({{\bf r},\tau,\lambda})\big]\,.
\end{equation}
Again we emphasize that in Eq. (\ref{F0}) the trace is taken over the $2\times 2$ matrices only.

In what follows we derive the first three terms in the expansion in Eqs. (\ref{F}, \ref{F0}) and obtain $S= S_0 + S_1 + S_2 +  S_{norm}$.

\subsection{The expansion of the action for a generic $\Delta (\tau, {\bf r})$}

We start with Eq. (\ref{F}).  Substituting Eqs. (\ref{Delta}, \ref{G}) into Eq. (\ref{ng}), we find that the zero's order Green's function ${\hat {\bf G}}_\lambda^{(0)} ({\bf r}, \tau-\tau^\prime; {\bf r}^\prime, \tau^\prime)$ satisfies the operator identity
\begin{equation}\label{ng0}
\big[
-\partial_\tau - {\cal H} ({\bf r},\tau',\lambda)
\big]{\hat {\bf G}}_\lambda^{(0)} ({\bf r}, \tau; {\bf r}^\prime, \tau^\prime) = \delta({\bf r}-{\bf r}^\prime)\delta(\tau-\tau^\prime)\,,
\end{equation}
where
\beq
{\cal H} ({\bf r},\tau',\lambda)  = {\hat K}({\bf r}) + {\hat \Delta}({\bf r},\tau',\lambda).
\eeq
The solution of this equation can be written as
\begin{equation}\label{G0}
{\hat {\bf G}}_\lambda^{(0)} ({\bf r}, \tau-\tau^\prime; {\bf r}^\prime, \tau^\prime) = \int {d\omega\over 2\pi} {\hat {\bf G}}_\lambda^{(0)} ({\bf r}, \omega; {\bf r}^\prime, \tau^\prime)\,e^{-i\omega (\tau-\tau^\prime)}\, ,
\end{equation}
with
\begin{equation}\label{G01}
{\hat {\bf G}}_\lambda^{(0)} ({\bf r}, \omega; {\bf r}^\prime, \tau^\prime) = \sum_n {|\chi_{n,\lambda}({\bf r},\tau')\rangle\langle \chi_{n,\lambda}({\bf r}^\prime,\tau')|\over i\omega - E_{n,\lambda}}    \, ,
\end{equation}
where
$|\chi_{n,\lambda}({\bf r},\tau')\rangle$ are the eigenfunctions of the corresponding Bogolubov-De Gennes equation:
\begin{equation}\label{bdg}
{\cal H} ({\bf r},\tau',\lambda)  |\chi_{n,\lambda}({\bf r},\tau')\rangle  =
E_{n,\lambda} (\tau')
|\chi_{n,\lambda}({\bf r},\tau')\rangle  \,,
\end{equation}
which satisfy  the completeness relation
\beq
\sum_n|\chi_{n,\lambda} ({\bf r})\rangle\langle \chi_{n,\lambda}  ({\bf r}')|=\delta({\bf r}-{\bf r}').
\label{short_1}
\eeq

The eigenfunctions $|\chi_{n,\lambda}({\bf r},\tau')\rangle$ depend parametrically  on $\lambda$ and $\tau'$. Continuing with the expansion, we find higher order contributions  in Eq. (\ref{G}) to be
\begin{equation}\label{G1}
 {\hat {\bf G}}_\lambda^{(1)} ({\bf r}, \omega; {\bf r}^\prime, \tau^\prime) =
i\int d{\bf r}_1\, {\hat {\bf G}}_\lambda^{(0)} ({\bf r}, \omega; {\bf r}_1, \tau^\prime) \big[\partial_{\tau^\prime}{\hat \Delta}({\bf r}_1,\tau^\prime,\lambda)\big]{\partial\over\partial\omega}{\hat {\bf G}}_\lambda^{(0)} ({\bf r}_1, \omega; {\bf r}^\prime, \tau^\prime)
\end{equation}

\begin{eqnarray}\label{G2}
{\hat {\bf G}}_\lambda^{(2)} ({\bf r}, \omega; {\bf r}^\prime, \tau^\prime) = -{1\over 2} \int d{\bf r}_1 \,{\hat {\bf G}}_\lambda^{(0)} ({\bf r}, \omega; {\bf r}_1, \tau^\prime) \big[\partial^2_{\tau^\prime}{\hat \Delta}({\bf r}_1,\tau^\prime,\lambda)\big]{\partial^2\over\partial\omega^2}{\hat {\bf G}}_\lambda^{(0)} ({\bf r}_1, \omega; {\bf r}^\prime, \tau^\prime)\ \ \ \ \ \ \ \ \ \ \\\nonumber
-\lambda^2 \int d{\bf r}_1 d{\bf r}_2\,{\partial\over\partial\omega} {\hat {\bf G}}_\lambda^{(0)} ({\bf r}, \omega; {\bf r}_1, \tau^\prime) \big[\partial_{\tau^\prime}{\hat \Delta}({\bf r}_1,\tau^\prime,\lambda)\big]{\hat {\bf G}}_\lambda^{(0)} ({\bf r}_1, \omega; {\bf r}_2, \tau^\prime) \big[\partial_{\tau^\prime}{\hat \Delta}({\bf r}_2,\tau^\prime,\lambda)\big] {\partial\over\partial\omega} {\hat {\bf G}}_\lambda^{(0)} ({\bf r}_2, \omega; {\bf r}^\prime, \tau^\prime)      \,.
\end{eqnarray}

We now substitute Eqs. (\ref{G0}, \ref{G01}, \ref{G1}, \ref{G2}) into Eqn. (\ref{F0}). The zeroth order term gives
\begin{equation}\label{F0fin1}
L^{(0)}(\tau)=\int_0^1 d\lambda \int {d\omega\over 2\pi} e^{i\omega\epsilon^+} \sum_n {\langle \chi_{n,\lambda}|\partial_\lambda{\hat \Delta}(\tau)|\chi_{n,\lambda}\rangle \over i\omega - E_{n,\lambda}} =
\sum_n \int_0^1 d\lambda\, \langle \chi_{n,\lambda}|\partial_\lambda{\hat \Delta}(\tau)|\chi_{n,\lambda}\rangle
\theta(-E_{n,\lambda})  \,.
\end{equation}
where $\theta (x) = 1$ for $x >0$ and $\theta (x) =0$ for $x <0$. Here and below we use the notation
\beq
\langle \chi_{n,\lambda}|...|\chi_{n,\lambda}\rangle  = \int d {\bf r} \langle \chi_{n,\lambda} ({\bf r})|...|\chi_{n,\lambda} ({\bf r})\rangle .
\label{short}
 \eeq

The integral over $\lambda$  in (\ref{F0fin1}) can be evaluated if we note that $
\partial_\lambda
{\hat \Delta}(\tau) = \partial_\lambda {\hat H}$.
Then
\begin{equation}\label{hf}
\langle \chi_{n,\lambda}|\partial_\lambda{\hat \Delta}(\tau)|\chi_{n,\lambda}\rangle = \langle \chi_{n,\lambda}|\partial_\lambda {\hat H}|\chi_{n,\lambda}\rangle = \partial_\lambda \langle \chi_{n,\lambda}|{\hat H}|\chi_{n,\lambda}\rangle = \partial_\lambda E_{n,\lambda}  \,.
\end{equation}
Substituting this into  (\ref{F0fin1}), integrating over $\lambda$,  and substituting the result into (\ref{F}),  we obtain  the zeroth order (adiabatic) term in the expansion of ${\cal S}$:
\begin{equation}\label{S0}
{\cal S}_0=\sum_n \int_{-\infty}^{\infty} d\tau \,\Big[E_n(\tau) -E_n^{(|\Delta| \to 0)}(\tau)\Big]\theta[-E_n(\tau)] +  \int_{-\infty}^{\infty} d\tau  \int d{\bf r} \frac{|\Delta ({\bf r},\tau)|^2}{g} \,.
\end{equation}
$E_n(\tau)$ in this  expression  are the eigen-energies $E_n(\tau)$ of the Bogolubov-De Gennes equation (\ref{bdg}) with $\lambda=1$.  (Note that $\sum_n E_n(\tau)$ is proportional to the area $S$ of a 2d system, so both terms in (\ref{S0}) scale as $S$.)
The counter-term  with $E_n^{(|\Delta| \to 0)}$ comes from the lower limit of the integration over $\lambda$.

To derive the term in the action with the first derivative over time, ${\cal S}_1 = \int d \tau L^{(1)}(\tau)$, we substitute Eq. (\ref{G1}) into Eq. (\ref{F0}). Evaluating the trace with the use of (\ref{short_1}), we obtain
\begin{eqnarray}\label{F1fin1}
{\cal S}_1 = i \int d \tau  \int_0^1 d\lambda \int {d\omega\over 2\pi} \sum_{n,m} {\langle \chi_{n,\lambda}|{\partial_\tau\hat \Delta}|\chi_{m,\lambda}\rangle \over i\omega - E_{n,\lambda}} {\partial\over\partial\omega} {\langle \chi_{m,\lambda}|\partial_\lambda{\hat \Delta}(\tau)|\chi_{n,\lambda}\rangle \over i\omega - E_{m,\lambda}} \\\nonumber
= \int d \tau \int_0^1 d\lambda \sum_{n,m} {\langle \chi_{n,\lambda}|{\partial_\tau\hat \Delta}|\chi_{m,\lambda}\rangle\langle \chi_{m,\lambda}|\partial_\lambda {\hat \Delta}|\chi_{n,\lambda}\rangle (\theta_{n,\lambda} -\theta_{m,\lambda}) \over (E_{n,\lambda} - E_{m,\lambda})^2}  \,,
\end{eqnarray}
We used Eq. (\ref{short}) and  a shorthand notation $\theta_{n,\lambda}\equiv \theta (E_{n,\lambda})$.
To proceed further, we use the identities
\begin{equation}\label{hfoff}
\langle \chi_{n,\lambda}|\partial_\lambda{\hat \Delta}(\tau)|\chi_{m,\lambda}\rangle = \langle \chi_{n,\lambda}|\partial_\lambda {\hat H}|\chi_{m,\lambda}\rangle =  \langle\partial_\lambda \chi_{n,\lambda}|\chi_{m,\lambda}\rangle (E_{n,\lambda} - E_{m,\lambda}) \ (m\neq n),
\end{equation}
\begin{equation}\label{hfofftau}
\langle \chi_{n,\lambda}|\partial_\tau{\hat \Delta}(\tau)|\chi_{m,\lambda}\rangle = \langle \chi_{n,\lambda}|\partial_\tau {\hat H}|\chi_{m,\lambda}\rangle =  \langle\partial_\tau \chi_{n,\lambda}|\chi_{m,\lambda}\rangle (E_{n,\lambda} - E_{m,\lambda}) \ (m\neq n).
\end{equation}
Substituting them into (\ref{F1fin1}), we get rid of the denominator. Integrating the rest by parts we obtain
\begin{eqnarray}\label{F1fin2}
{\cal S}_1 =
\int d \tau \int_0^1 d\lambda \sum_n \Big[\langle \partial_\lambda\chi_{n,\lambda}|\partial_\tau\chi_{n,\lambda}\rangle - \langle \partial_\tau\chi_{n,\lambda}|\partial_\lambda\chi_{n,\lambda}\rangle \Big] \theta_{n,\lambda} \,.
\end{eqnarray}
One can make sure~\cite{com2} that the term inside the brackets is
\beq
\frac{d}{d\lambda}  [\langle\chi_{n}(\tau)|\partial_\tau\chi_{n}(\tau)\rangle - \langle\chi_{n}^{(|\Delta| \to 0)}(\tau)|\partial_\tau\chi_{n}^{(|\Delta| \to 0)}(\tau)\rangle].
\eeq
Then
\begin{equation}\label{S1}
{\cal S}_1=\sum_n \int d\tau\, \Big[\langle\chi_{n}(\tau)|\partial_\tau\chi_{n}(\tau)\rangle - \langle\chi_{n}^{(|\Delta| \to 0)}(\tau)|\partial_\tau\chi_{n}^{(|\Delta| \to 0)}(\tau)\rangle\Big]\theta_{n,\lambda}
\end{equation}

The derivation of the second order  term in the action $S_2 = \int d \tau L^{(2)} (\tau)$ is more cumbersome. We present the details in Appendix A. The result is
\begin{equation}\label{S2}
{\cal S}_2= -{1\over 2}\,\sum_{m\neq n}
\int  d\tau \Big[ {|\langle\chi_{n}(\tau)|\partial_\tau\chi_{m}(\tau)\rangle|^2(\theta_n-\theta_m)\over E_n(\tau)-E_m(\tau)} - {|\langle\chi_{n}^{(|\Delta| \to 0)}(\tau)|\partial_\tau\chi_{m}^{(|\Delta| \to 0)}(\tau)\rangle|^2(\theta_n-\theta_m)\over E_n^{(|\Delta| \to 0)}(\tau)-E_m^{(|\Delta| \to 0)}(\tau)} \Big]   \,.
\end{equation}

We emphasize that in Eqs. (\ref{S1}) and (\ref{S2}), the wave-functions $|\partial_\tau\chi_{n,\lambda}(\tau)\rangle$ and energies $E_n(\tau)$ satisfy Bogoluibov-De Gennes equations, in which the order parameter $\Delta({\bf r},\tau)$ depends on coordinate ${\bf r}$
and on $\tau$.

Finally, consider ${\cal S}_{norm} = - Tr \log{{\hat G}^{-1}_{\Delta \to 0}}$,  Eq. (\ref{chn_1}).  We argue that it also contains the term linear in time derivative. The most straightforward way to see this is to keep $\Delta$ small but finite and apply a gauge transformation under the logarithm to get rid of the $\phi$ dependence of $\Delta (\phi) = \Delta e^{i \phi}$, i.e.,  replace ${\hat G}^{-1}_{\Delta \to 0}$ by ${\hat U}^\dagger {\hat G}^{-1}_{\Delta \to 0} {\hat U} $, where ${\hat U}$ is chosen such that in ${\hat U}^\dagger{\hat G}^{-1}_{\Delta \to 0}  {\hat U}$, $\Delta$ appears without $e^{i\phi}$ factor (Ref. \cite{thouless,ao_1}). A simple experimentation shows that one should choose ${\hat U}$ in the form
\begin{equation}
\label{ss_1}
{\hat U}({\bf r},\tau) =
\begin{bmatrix}
e^{i\phi ({\bf r}, \tau)/2} & 0 \\
 0 & e^{-i\phi ({\bf r}, \tau)/2}
\end{bmatrix}\,.
\end{equation}
Once $\Delta (\phi)$ is stripped of the phase, its magnitude can be safely set to zero.  However, because  ${\hat G}^{-1}$ contains time and spatial derivatives, ${\hat U}^\dagger{\hat G}^{-1}_{\Delta \to 0}  {\hat U} $ acquires the terms with ${\dot \phi}$ and $\nabla \phi$.  These terms are additional to the ones in $S_1$ because to obtain  the latter we used the expansion in powers of $\Delta$, while here we treat $\Delta$ as infinitesimally small and do not expand in it.

Using (\ref{ng}) and (\ref{ss_1}),  we obtain, keeping only time derivative of $\phi$,
\begin{equation}
\label{ss_211}
{\cal S}_{norm} =  - Tr \log{\big[{\hat G}^{-1}_0 -\frac{i}{2} \sigma_z {\dot \phi} ({\bf r}, \tau)\big]} \nonumber \\
\end{equation}
where
\begin{equation}
\label{ss_21}
{\hat G}_0^{-1} =
\begin{bmatrix}
-\partial_\tau+(1/2m)(\nabla+(i/2)\nabla\phi)^2+\mu & 0 \\
0 & -\partial_\tau-(1/2m)(\nabla-(i/2)\nabla\phi)^2-\mu
\end{bmatrix}\,.
\end{equation}

${\cal S}_{norm}$ in Eqs. (\ref{ss_211}), (\ref{ss_21}), as well as ${\cal S}_0$, ${\cal S}_1$, ${\cal S}_2$ can be expanded in terms of space and time derivatives of $\phi$.  This will be carried out in the next subsection, where we will analyze the long wavelength - low frequency limit of the effective action derived in this subsection and obtain the Anderson-Bogolubov-Goldstone (ABG) mode of gapless phase fluctuations. A special attention is required when the phase $\phi$ contains a vortex, in which case an expansion in $\nabla\phi$ fails in the vicinity of the vortex core. Instead we expand of ${\cal S}_{norm}$ in terms of small displacements of the vortex core and show that there is a transverse reaction force associated with the readjustments of the normal component to the vortex displacement. The corresponding analysis will be carried out in Sec. IV.

\subsection{The long wavelength limit}

The expansion of the action in the previous section holds for any function $\Delta (\tau, {\bf r})$.
In this section we consider slowly varying order parameter and derive an effective action in terms of its spatial and time derivatives. We compute each term in $S= S_0 + S_1 + S_2 +  S_{norm}$ separately

\subsubsection{The action ${\cal S}_0$: the condensation energy and the $(\nabla \phi)^2$ term.}

The term ${\cal S}_0$ is given by Eq. (\ref{S0}). To express it in terms of spatial derivatives of $\Delta$, we need to find the solutions to the Bogolubov-de Gennes equation:
\begin{equation}\label{bdgmain}
\begin{bmatrix}
-(1/2m)\nabla^2-\mu & |\Delta({\bf r},\tau)|e^{i\phi({\bf r},\tau)} \\
|\Delta({\bf r},\tau)|e^{-i\phi({\bf r},\tau)} & (1/2m)\nabla^2+\mu
\end{bmatrix}
\begin{bmatrix}
u_n({\bf r},\tau) \\
v_n({\bf r},\tau)
\end{bmatrix}
=E_n(\tau)
\begin{bmatrix}
u_n({\bf r},\tau) \\
v_n({\bf r},\tau)
\end{bmatrix}\,,
\end{equation}
To get rid of the complex phase of $\Delta({\bf r},\tau)$, we redefine the wavefunction in Eq. (\ref{bdgmain}) as $|\chi_n({\bf r},\tau)\rangle = e^{i\phi({\bf r},\tau) \sigma_z/2}|{\tilde\chi}_n({\bf r},\tau)\rangle$. The eigenfunction $|{\tilde\chi}_n({\bf r},\tau)\rangle$ satisfies the equation
\begin{equation}\label{ham}
\bigg\{{\hat H}_0 - {i\over 4m}\big[\nabla(\nabla\phi)+(\nabla\phi)\nabla\big]+{{\hat\sigma}_z\over 8m}(\nabla\phi)^2\bigg\}|{\tilde\chi}_n({\bf r},\tau)\rangle =E_n (\tau) |{\tilde\chi}_n({\bf r},\tau)\rangle     \,.
\end{equation}
where
\begin{equation}\label{h0}
{\hat H}_0 =
\begin{bmatrix}
-(1/2m)\nabla^2-\mu & |\Delta({\bf r},\tau)| \\
|\Delta({\bf r},\tau)| & (1/2m)\nabla^2+\mu
\end{bmatrix}\ .
\end{equation}
Due to slow variation of $\phi$ on ${\bf r}$,  the last two terms in Eq. (\ref{ham}) can be treated as perturbations. We label then as ${\hat V}_1$ and ${\hat V}_2$:
\begin{equation}\label{V}
{\hat V}_1 = - {i\over 4m}\big[\nabla(\nabla\phi)+(\nabla\phi)\nabla\big] \ ,\ {\hat V}_2={{\hat\sigma}_z\over 8m}(\nabla\phi)^2
\end{equation}
In  ${\hat V}_1$ the free-standing gradient in the first term acts on the bra state on the left, and in the second term it acts on the ket state on the right.

In the following we restrict our analysis to terms quadratic in spatial derivatives.  It is easy to verify that to this order once can neglect the spatial fluctuations of $|\Delta({\bf r},\tau)|$ as the spatial dependence of $|\Delta|$ only gives rise to  third order terms like  $(\nabla\phi)^2 (\nabla \Delta)$, etc..

The  eigen-states of (\ref{ham}) at ${\hat V}_1 ={\hat V}_2 =0$ are the conventional Bogolubov solutions, for which $n$ is a continuous 2d variable, which we label as ${\bf k}$
For the particle branch we have
\begin{equation}\label{chi0+}
|{\tilde\chi}_{\bf k}^{(+)}({\bf r})\rangle \equiv
\begin{bmatrix}
{\tilde u}_{\bf k} \\
{\tilde v}_{\bf k}
\end{bmatrix}
e^{i{\bf kr}}
=
\begin{bmatrix}
\sqrt{{1\over 2}+{\xi_{\bf k}\over 2|E_{\bf k}|}} \\
\sqrt{{1\over 2}-{\xi_{\bf k}\over 2|E_{\bf k}|}}
\end{bmatrix}
{e^{i{\bf kr}}\over \sqrt{S}}\ ,
\end{equation}
where $E_n (\tau)= E_{\bf k}^{(+)}(\tau)=+\sqrt{\xi_{\bf k}^2+|\Delta(\tau)|^2}$ and, we remind,  $S$ is the area of the sample. For the hole branch we have
\begin{equation}\label{chi0-}
|{\tilde\chi}_{\bf k}^{(-)}({\bf r})\rangle \equiv
\begin{bmatrix}
{\tilde v}_{\bf k} \\
-{\tilde u}_{\bf k}
\end{bmatrix}
e^{i{\bf kr}}
=
\begin{bmatrix}
\sqrt{{1\over 2}-{\xi_{\bf k}\over 2|E_{\bf k}|}} \\
-\sqrt{{1\over 2}+{\xi_{\bf k}\over 2|E_{\bf k}|}}
\end{bmatrix}
{e^{i{\bf kr}}\over \sqrt{S}} \ ,
\end{equation}
where  $E_n(\tau) = E_{\bf k}^{(-)}(\tau)=-\sqrt{\xi_{\bf k}^2+|\Delta(\tau)|^2}$.\\

\paragraph{The condensation energy:\\}

We label by ${\cal S}_{0,a}$ the  term in ${\cal S}_0$, which does not contain gradients. It is given by
\begin{equation}\label{E0}
{\cal S}_{0,a} (\Delta) = \int d \tau \int d{\bf r} \left[-  \int {d^2{\bf k}\over (2\pi)^2}\,\Big\{\sqrt{\xi_{\bf k}^2+|\Delta(\tau)|^2} - |\xi_{\bf k}|\Big\} +
\frac{|\Delta (\tau)|^2}{g} \right]\,.
\end{equation}

In equilibrium, $\Delta (\tau)=\Delta_0 +\delta\Delta (\tau)$, where $\Delta_0\gg\delta\Delta (\tau)$. Substituting this $\Delta (\tau)$ into (\ref{E0}) and using $\partial {\cal S}_{0,a} (\Delta) /\partial (\delta \Delta) =0$, we obtain a conventional gap equation
\begin{equation}\label{self}
S {\Delta_0\over g}  = \Delta_0 \sum_{\omega,{\bf k}} {\Delta_0\over \omega^2 + \Delta_0^2 + \xi^2_{\bf k}}\,,
\end{equation}
which after the integration over  Matsubara frequency becomes
\beq
\frac{1}{g} = \frac{N_0}{2} \int_{-\mu}^\Lambda \frac{d \xi}{\sqrt{\xi^2+\Delta^2_0}}
\label{bb2_2}
\eeq
Integrating further over $\xi$ in (\ref{bb2_2}) and re-expressing the result in terms of the bound state energy $2E_0 = 2\Lambda e^{-2/(N_0g)}$, we obtain the relation~\cite{Miyake1983,Mohit89,chubukov_17}
\begin{equation}\label{Delta0}
\sqrt{\mu^2 + \Delta_0^2}-\mu = 2E_0  \,,
\end{equation}
The self-consistency equation for $\mu$ in turn follows from the  condition that the total number of fermions, including bound pairs, is conserved~\cite{Mohit89}. This  gives another relation
\begin{equation}\label{Delta1}
\sqrt{\mu^2 + \Delta_0^2}+\mu = 2E_F  \,.
\end{equation}
 Solving Eqs. (\ref{Delta0}, \ref{Delta1}) we obtain
\begin{equation}\label{Delta_mu}
 \mu = E_F - E_0\,,~~~ \Delta_0 = 2 \sqrt{E_F E_0}.
\end{equation}
We will use these formulas below when we evaluate the prefactors for ${\dot \phi}$, $(\dot \phi)^2$ and $(\nabla \phi)^2$ terms in the crossover region between BCS and BEC behavior.  We recall that  BCS behavior holds when the bound state energy $E_0$ is much smaller than $E_F$ (and $\Delta_0 \ll \mu$)  and BEC behavior holds when $E_0 \gg E_F$.  A negative $\mu$ at $E_F < E_0$ implies that  the Fermi momentum $k_F$, defined as position of the minimum of  the fermionic dispersion $E_k = \sqrt{(\varepsilon_k - \mu)^2 + \Delta^2_0}$), is zero~\cite{Mohit2015}.

Eqs. (\ref{Delta1}) and (\ref{Delta_mu}) allow one to obtain the condensation energy of a superconductor in the whole crossover range between BCS and BEC regimes.  We have
\beq
E_{cond} - NS (\mu - \mu_0)  = {\cal S}_{0,a} + \delta {\cal S}_{free}
\eeq
where $\delta {\cal S}_{free}$ is the difference between $2 \sum_k \xi_k n_k $ in the normal state at chemical potentials $\mu$ and $\mu_0$.
Using $N = 2 N_0 E_F, \mu-\mu_0 = -E_0$ and evaluating $\delta {\cal S}_{free} = S N_0 (\mu^2_0 - \mu^2)$ for $\mu >0$ and $\delta {\cal S}_{free} = S N_0 \mu^2_0$ for $\mu <0$,  we obtain
\bea
E_{cond} &=& - S N_0 E^2_0 + {\cal S}_{0,a},   ~~\mu >0 \nonumber \\
&=& - S N_0 \left(\frac{\Delta^2}{2} - E^2_F\right) +  {\cal S}_{0,a} ~~\mu <0
\label{t1}
\eea
Using (\ref{bb2_2}) and introducing $x = \mu/\Delta_0$ and $y = \xi/\Delta_0$, we re-express $S_{0,a}$ in (\ref{E0}) as
\beq
{\cal S}_{0,a} = S N_0 \frac{\Delta^2_0}{2} I(x)
\eeq
where
\beq
I(x) = \int^{\infty}_{-x} dy \left(\frac{1}{\sqrt{y^2+1}} -2\sqrt{y^2+1} + |y|\right)
\eeq
This integration yields
\bea
I(x) &=& -\frac{1}{2} + x^2 -x \sqrt{1+x^2}, ~~ x >0 \nonumber \\
&=& -\frac{1}{2} -x^2 + |x| \sqrt{1+x^2}, ~~ x<0
\eea
Substituting the expressions for $\mu$ and $\Delta$ we obtain
\bea
{\cal S}_{0,a} &=&  S N_0 \left(-\frac{\Delta^2_0}{2} + E^2_0 \right),   ~~\mu >0, \nonumber\\
&=& - S N_0 E^2_F,  ~~ \mu <0
\label{t2}
\eea
The combination of (\ref{t1}) and (\ref{t2}) yields
\beq
E_{cond} = -S N_0 \frac{\Delta^2_0}{2}
\label{t3}
\eeq
independent on the ratio $E_0/E_F$.  The same result (the independence of $ E_{cond}$ on $E_0/E_F$) has been  also obtained~\cite{chubukov_17,we_1} by directly evaluating the kinetic and the potential energy of a superconductor.\\

\paragraph{The $\nabla\phi$ term:\\}

The  leading term in  $\nabla\phi$ in ${\cal S}_0$  can be calculated by treating $\nabla\phi$ terms in the Hamiltonian in Eq. (\ref{ham}) as perturbations. To the first order this contribution is given by ${\hat V}_2$. We label the corresponding term in ${\cal S}_0$ as ${\cal S}_{0,b}$. We have
\begin{equation}\label{phi1}
{\cal S}_{0,b} = -\int d \tau \sum_{\bf k}\left[\langle{\tilde\chi}_{\bf k}^{(-)}|{\hat V}_2|{\tilde\chi}_{\bf k}^{(-)}\rangle_{\Delta} - \langle{\tilde\chi}_{\bf k}^{(-)}|{\hat V}_2|{\tilde\chi}_{\bf k}^{(-)}\rangle_{\Delta \to 0}\right]
\theta (-E_k)
 = - \Big[\frac{1}{S} \sum_{\bf k}\big({\xi_{\bf k}\over |E_{\bf k}|}-{\xi_{\bf k}\over |\xi_{\bf k}|}\big)\Big] \int d{\bf r}
{(\nabla\phi)^2\over 8m} \,.
\end{equation}
The k-integral is ultraviolet convergent. Note that due to the presence of $\theta(-E_n)$ in Eq. (\ref{S0}), the summation in Eq. (\ref{phi1}) involves only the hole states $|{\tilde\chi}_{\bf k}^{(-)}\rangle$.  Evaluating the integral we find
\begin{equation}\label{sumb}
- \frac{1}{S} \sum_{\bf k}\big({\xi_{\bf k}\over |E_{\bf k}|}-{\xi_{\bf k}\over |\xi_{\bf k}|}\big) =
- \int {d^2{\bf k}\over (2\pi)^2}\big({\xi_{\bf k}\over |E_{\bf k}|}-{\xi_{\bf k}\over |\xi_{\bf k}|}\big)= N_0\big(\sqrt{\mu^2+|\Delta (\tau)|^2} - |\mu|\big)     \,.
\end{equation}
This term can be equivalently re-expressed as
\beq
\int {d^2{\bf k}\over (2\pi)^2}\big(1 -{\xi_{\bf k}\over |E_{\bf k}|}\big) -  \int {d^2{\bf k}\over (2\pi)^2} \big(1-{\xi_{\bf k}\over |\xi_{\bf k}|}\big) =
2 \int {d^2{\bf k}\over (2\pi)^2} \left( ({\tilde v}^2_k)_{\Delta} - {\tilde v}^2_k (\Delta=0) \right) = n-n_0
\label{bb5}
\eeq
where
\beq
n = 2 \int {d^2{\bf k}\over (2\pi)^2} ({\tilde v}^2_k)_{\Delta} =  \big(\sqrt{\mu^2+|\Delta (\tau)|^2} + |\mu|\big) =
2 N_0 E_F
\eeq
is the density of fermions and
\beq
n_0 = 2 \int {d^2{\bf k}\over (2\pi)^2} ({\tilde v}^2_k)_{\Delta =0} = 2N_0 \mu \theta(\mu)
\eeq
is the density of free electrons in the normal state at the actual chemical potential $\mu$.
Using these notations, we find
\beq
{\cal S}_{0,b} =  (n-n_0) \int d{\bf r} {(\nabla \phi)^2\over 8m}
\label{e_2_a}
\eeq

The last contribution to ${\cal S}_0$ comes from  ${\hat V}_1$. The first order contribution from ${\hat V}_1$ is zero. The contribution to order $({\hat V}_1)^2$ is given by
\begin{equation}
{\cal S}_{0,c}  = ({\cal S}_{0,c})_{\Delta} - ({\cal S}_{0,c})_{\Delta=0}
\label{bb_9}
\eeq
where
\beq
\label{phi2}
({\cal S}_{0,c})_{\Delta} = \sum_{{\bf k},{\bf k}^\prime, i=\pm}{{\langle{\tilde\chi}_{\bf k}^{(-)}|{\hat V}_1|{\tilde\chi}_{{\bf k}^\prime}^{(i)}\rangle}
\langle{\tilde\chi}_{{\bf k}^\prime}^{(i)}|{\hat V}_1|{\tilde\chi}_{\bf k}^{(-)}\rangle \over E_{\bf k}^{(-)}-E_{{\bf k}^\prime}^{(i)}}\, = ({\cal S}^{-+}_{0,c})_{\Delta} +
({\cal S}^{--}_{0,c})_{\Delta}.
\end{equation}
Consider $({\cal S}^{-+}_{0,c})_{\Delta}$ and $({\cal S}^{--}_{0,c})_{\Delta}$ separately.  For $i =+$, we use Eq. (\ref{chi0+}) for
$|{\tilde\chi}_{\bf k}^{(+)}({\bf r})\rangle$ and Eq. (\ref{chi0-}) for
$|{\tilde\chi}_{\bf k}^{(-)}({\bf r})\rangle$ and obtain
\begin{equation}\label{prod1}
{\langle{\tilde\chi}_{\bf k}^{(-)}|{\hat V}_1|{\tilde\chi}_{{\bf k}^\prime}^{(+)}\rangle}
\langle{\tilde\chi}_{{\bf k}^\prime}^{(+)}|{\hat V}_1|{\tilde\chi}_{\bf k}^{(-)}\rangle = ({\tilde u}_{\bf k}{\tilde v}_{{\bf k}^\prime} -{\tilde v}_{\bf k}{\tilde u}_{{\bf k}^\prime})^2 {(k_j+k_j^\prime)^2\over 2m }  \int d{\bf r}d{\bf r}^\prime {(\nabla_j\phi)(\nabla_{j}^\prime\phi)\over 8m} e^{i({\bf k}-{\bf k}^\prime)({\bf r}-{\bf r}^\prime)}  \,.
\end{equation}
where $j=x,y$.
Using the forms of ${\tilde u}_k$ and ${\tilde v}_k$, we then obtain
\begin{equation}\label{num1}
({\cal S}^{-+}_{0,c})_{\Delta} = - {1\over S^2}\sum_{{\bf k},{\bf k}^\prime}\ {E_{\bf k}E_{{\bf k}^\prime}-\xi_{\bf k}\xi_{{\bf k}^\prime}-\Delta_0^2 \over 2 E_{\bf k}E_{{\bf k}^\prime} (E_{\bf k}+E_{{\bf k}^\prime})}\ {(k_j+k_j^\prime)^2\over 2m }\ \int d{\bf r}d{\bf r}^\prime {(\nabla_j\phi)(\nabla_j^\prime\phi)\over 8m} e^{i({\bf k}-{\bf k}^\prime)({\bf r}-{\bf r}^\prime)} \,,
\end{equation}
where in the prefactor we can use the zero-order expression $E_{\bf k}= \sqrt{\xi^2_k + \Delta^2_0}$.
For $({\cal S}^{--}_{0,c})_{\Delta}$ the computation  along the same lines yields
\begin{equation}\label{num2}
({\cal S}^{--}_{0,c})_{\Delta} = -
{1\over S^2}\sum_{{\bf k},{\bf k}^\prime}\ {E_{\bf k}E_{{\bf k}^\prime}+\xi_{\bf k}\xi_{{\bf k}^\prime}+\Delta_0^2 \over 2 E_{\bf k}E_{{\bf k}^\prime} (E_{\bf k}-E_{{\bf k}^\prime})}\ {(k_j+k_j^\prime)^2\over 2m }\ \int d{\bf r}d{\bf r}^\prime {(\nabla_j\phi)(\nabla_j^\prime\phi)\over 8m} e^{i({\bf k}-{\bf k}^\prime)({\bf r}-{\bf r}^\prime)} \,.
\end{equation}
The total contribution $({\cal S}_{0,c})_{\Delta} = ({\cal S}^{-+}_{0,c})_{\Delta} + ({\cal S}^{--}_{0,c})_{\Delta}$ is, after symmetrization over ${\bf k}$ and ${\bf k}'$
\begin{equation}\label{numtot}
({\cal S}_{0,c})_{\Delta} = - {1\over 2 S^2}\sum_{{\bf k},{\bf k}^\prime}\ {E_{\bf k} E_{{\bf k}'}
-(\xi_{\bf k}\xi_{{\bf k}^\prime}+\Delta_0^2)\over E_{\bf k}E_{{\bf k}'} (E_{{\bf k}} + E_{{\bf k}^\prime})}\ {(k_j+k_j^\prime)^2\over 2m }\ \int d{\bf r}d{\bf r}^\prime {(\nabla_j\phi)(\nabla_j^\prime\phi)\over 8m} e^{i({\bf k}-{\bf k}^\prime)({\bf r}-{\bf r}^\prime)} \,.
\end{equation}

To proceed, assume that $(\nabla_j\phi ({\bf r}))(\nabla_j^\prime\phi ({\bf r}^\prime))$, viewed as a function of $\delta {\bf r} = {\bf r}-{\bf r}^\prime$, drops at some characteristic scale $D_0$, which is much smaller than the system size $2D$, but much larger than interatomic spacing $a_0$.  The corresponding characteristic $\delta_{\bf k} = |{\bf k}-{\bf k}^\prime|$ are of order  $1/D_0$, which satisfies $1/D \ll 1/D_0 \ll 1/a_0$.  Such  $\delta k$ are, on one hand, much smaller than $k_F$, and, on the other hand, are large enough such that the discreteness of momentum $\delta k_m = \pi m/D$ is irrelevant. As the consequence, the expression for $({\cal S}_{0,c})_{\Delta}$ can be re-expressed, to leading order in the derivatives, as
\begin{equation}\label{bb_6}
({\cal S}_{0,c})_{\Delta} = -  {\text lim}_{\delta {\bf k} \to 0} \chi_j (\delta k)  \int d{\bf r} {(\nabla_j\phi)^2\over 8m},
\end{equation}
where we introduced
\beq
\chi_j (\delta {\bf k}) =~{1\over S} \sum_{{\bf k}}\ {E_{\bf k_{-}} E_{{\bf k}_{+}}
-(\xi_{{\bf k}_{+}}\xi_{{\bf k}_{-}}+\Delta_0^2
)\over E_{{\bf k}_+}  E_{{\bf k}_{-}} (E_{{\bf k}_+} + E_{{\bf k}_-})}\ \frac{k^2_j}{m}\
\label{bb_7}
\eeq
with  ${\bf k}_{\pm} = {\bf k} \pm \delta{\bf k}/2$, and used
\beq
\int \frac{d (\delta k_j)}{2\pi} \int_{-D_0}^{D_0} d ({\delta r}_j)  e^{i \delta k_j \delta r_j} = \frac{2}{\pi} \int_0^\infty \frac{sin{x}}{x} dx =1.
\eeq

The quantity $\chi (\delta {\bf k})$ is, up an overall factor, a particle-hole bubble made out of superconducting Green's functions.
At a finite  $\Delta$,  it vanishes at ${\delta {\bf k} \to 0}$ because the term in the numerator in  (\ref{bb_7}) tends to zero in this limit.
Accordingly, $({\cal S}_{0,c})_{\Delta} =0$. However, for $({\cal S}_{0,c})_{\Delta \to 0}$, the corresponding $\chi (\delta {\bf k})$ is a free-fermion static susceptibility in the normal state, and it tends to a finite value when $\delta {\bf k}$  is small but finite.  We now use the fact that at small $\delta {\bf k}$ the integration over ${\bf k}$  in (\ref{bb_7}) is confined to ${\bf k} = {\bf k}_F$ and pull $k^2_j/m \approx (k^2_F)_j/m$ from the sum. Performing the remaining integration with
$E_{\bf k} = |\xi_{{\bf k}}|$ and using the symmetry between $j=x$ and $j=y$ and the fact that $N_0 k^2_F/m = 2N_0 \mu \theta (\mu) = n_0$,  we obtain
\beq
({\cal S}_{0,c})_{\Delta \to 0} = - N_0 \frac{k^2_F}{m}  \int d{\bf r} {(\nabla \phi)^2\over 8m} = - n_0 \int d{\bf r} {(\nabla \phi)^2\over 8m}
\label{bb_8}
\eeq
Substituting this into (\ref{bb_9}), we obtain
\beq
{\cal S}_{0,c} =  n_0 \int d{\bf r} {(\nabla \phi)^2\over 8m}
\label{e_2}
\eeq
Combining (\ref{e_2_a}) and (\ref{e_2})   we obtain the total  term with $(\nabla\phi)$ and no time derivative in the form
\begin{equation}\label{totphifin}
{\cal S}_{0,b} +  {\cal S}_{0,c} =  n \int d\tau \int d{\bf r} \frac{(\nabla\phi)^2}{8m}\,.
\end{equation}
We see that the prefactor for the $(\nabla\phi)^2$ term in the action is the full density.  The consideration can be readily extended to  the case when impurity scattering is present.  The result is that $n$ is replaced by the superfluid density $n_s$.  In our consideration we do not distinguish between $n$ and $n_s$.

\subsubsection{The actions ${\cal S}_1$ and ${\cal S}_{norm}$ -- the linear term in ${\dot \phi}$. }

The calculation of the first order term in the derivative over $\tau$  is quite straightforward. We start with $S_1$ term.
From Eqs. (\ref{S1}),  (\ref{chi0-}) we obtain
\beq
\sum_n  \langle\chi_{n}(\tau)|\partial_\tau\chi_{n}(\tau)\rangle - \langle\chi_{n}^{(|\Delta| \to 0)}(\tau)|\partial_\tau\chi_{n}^{(|\Delta| \to 0)}(\tau)\rangle
\theta_{n,\lambda} =
\eeq
\beq
i \int d{\bf r}\,{\dot\phi({\bf r},\tau)\over 2} \frac{1}{S}\sum_{\bf k} \Big[({\tilde v}^2_{\bf k}-{\tilde u}^2_{\bf k})_{\Delta\neq 0} - ({\tilde v}^2_{\bf k}- {\tilde u}^2_{\bf k})_{\Delta = 0} \Big]\,
=i \int d{\bf r}\, {\dot \phi({\bf r},\tau)\over 2}\, \frac{1}{S} \sum_{\bf k} \Big({\xi_{\bf k}\over |\xi_{\bf k}|} -{\xi_{\bf k}\over |E_{\bf k}|}\Big)\,
\label{b1}
\eeq
Substituting this into  Eq. (\ref{S1}) and using  Eq. (\ref{sumb}),  we obtain
\begin{equation}\label{b2}
{\cal S}_1 =
\frac{iN_0}{2}\, \int d \tau \int d {\bf r}\,\Big(\sqrt{\mu^2+|\Delta(\tau)|^2}-|\mu|\Big)\, \dot \phi({\bf r},\tau) \, = i  \int d \tau \int d {\bf r}
\frac{ n(\tau) -n_0}{2}  \dot \phi({\bf r},\tau)
\end{equation}
Note that this expression again contains fluctuating $\Delta (\tau)$ rather than equilibrium $\Delta_0$.

Eq. (\ref{b2}) can be cast in the form of the Wess-Zumino action for a superconductor \cite{volovik,volovik_book,volovik_1}. To see this, let's recall the derivation of ${\cal S}_1$, e.g. Eq. (\ref{S1}), and write it as a slightly modified version of  Eq.(\ref{F1fin1}),
\begin{eqnarray}\label{WZ1}
{\cal S}_1=
\int_{-\infty}^{\infty}d\tau\int_0^1 d\lambda\, \sum_{n,m} {\langle \chi_{n,\lambda}|{\partial_\lambda\hat \Delta}|\chi_{m,\lambda}\rangle\langle \chi_{m,\lambda}|\partial_\tau{\hat \Delta}|\chi_{n,\lambda}\rangle (\theta_{n,\lambda} - \theta_{m,\lambda}) \over (E_{n,\lambda} - E_{m,\lambda})^2}  \,.
\end{eqnarray}

Since Eq. (\ref{WZ1}) already contains double gradients (over $\tau$ and $\lambda$) we can treat states $|\chi_{n,\lambda}\rangle$ and energies $E_{n,\lambda}$ in this equation adiabatically, i.e., use Eq. (\ref{chi0+}) and  $E_{\bf k,\lambda}^{(+)}(\tau)=+\sqrt{\xi_{\bf k}^2+|\Delta(\tau,\lambda)|^2}$ for the particle branch and use Eq. (\ref{chi0-}) and $E_{\bf k,\lambda}^{(-)}(\tau)=-E_{{\bf k},\lambda}^{(+)}(\tau)$ for the hole branch.

The integrand in Eq. (\ref{WZ1}) can then be written as
\begin{eqnarray}\label{WZ2}
{1\over S^2}\sum_{{\bf k},{\bf k}^\prime}{1 \over (E_{{\bf k},\lambda} + E_{{\bf k}^\prime,\lambda})^2}\int d{\bf r} d{\bf r}^\prime
e^{i({\bf k}-{\bf k}^\prime)({\bf r}-{\bf r}^\prime)}
\Big\{\big[\partial_\lambda \Delta({\bf r}) v_{{\bf k},\lambda} v_{{\bf k}^\prime,\lambda} - \partial_\lambda \Delta^\ast({\bf r}) u_{{\bf k},\lambda} u_{{\bf k}^\prime,\lambda} \big]\times\nonumber\\
\big[\partial_\tau \Delta^\ast({\bf r}^\prime) v_{{\bf k},\lambda} v_{{\bf k}^\prime,\lambda} - \partial_\tau \Delta({\bf r}^\prime) u_{{\bf k},\lambda} u_{{\bf k}^\prime,\lambda}\big]  -\big({\rm same\ with}\ \Delta\rightarrow\Delta^\ast\big) \Big\}\,.
\end{eqnarray}
In the long-wavelength limit we can replace $(1/S^{2})\sum_{{{\bf k},{\bf k}^\prime}}e^{i({\bf k}-{\bf k}^\prime)({\bf r}-{\bf r}^\prime)}v^2_{{\bf k},\lambda} v^2_{{\bf k}^\prime,\lambda}/(E_{{\bf k},\lambda} + E_{{\bf k}^\prime,\lambda})^2$ by $C_v\,\delta({\bf r}-{\bf r}^\prime)$ with
\begin{equation}\nonumber
C_v= {1\over S^2}\sum_{{\bf k},{\bf k}^\prime}\ {v^2_{{\bf k},\lambda} v^2_{{\bf k}^\prime,\lambda}\over (E_{{\bf k},\lambda} + E_{{\bf k}^\prime,\lambda})^2}\, (2\pi)^2\delta({\bf k}-{\bf k}^\prime) = {1\over S}\sum_{{\bf k}}\ {v^4_{{\bf k},\lambda}\over 4E^2_{{\bf k},\lambda} }   \,,
\end{equation}
etc. Then Eq. (\ref{WZ2}) reduces to a single integral over ${\bf r}$, which can be expressed as
\begin{equation}\label{WZ3}
{1\over S}\sum_{{\bf k}}\ {u^2_{{\bf k},\lambda}-v^2_{{\bf k},\lambda}\over 4E^2_{{\bf k},\lambda}}\int d{\bf r}\, \big(\partial_\lambda\Delta\partial_\tau\Delta^\ast - \partial_\tau\Delta\partial_\lambda\Delta^\ast \big)\,,
\end{equation}
where we have used that $v^2_{{\bf k},\lambda}+u^2_{{\bf k},\lambda}=1$. Finally, using
\begin{equation}\nonumber
{1\over S}\sum_{{\bf k}}\ {u^2_{{\bf k},\lambda}- v^2_{{\bf k},\lambda}\over 4E^2_{{\bf k},\lambda} } = {1\over 2}{\partial n\over \partial(|\Delta|^2)}   \,,
\end{equation}
where $n$ is particle density,
we express the action $S_1$ as
\begin{equation}\label{WZ4}
{\cal S}_1=
{1\over 2}\int d{\bf r} \int_{-\infty}^{\infty}d\tau\int_0^1 d\lambda\, {\partial n\over \partial(|\Delta|^2)} \big(\partial_\lambda\Delta\partial_\tau\Delta^\ast - \partial_\tau\Delta\partial_\lambda\Delta^\ast \big)\,.
\end{equation}
This action has the same form as Wess-Zumino action for s-wave superconductor \cite{volovik,volovik_book}. Note, however, that Eq. (\ref{WZ4}) is only valid in the long wave length limit, e.g., it does not account for the bound states that may arise in a vortex core \cite{caroli}, whereas Eq. (\ref{S1}) is more general because  it includes all types of  states.

We now turn to the contribution from ${\cal S}_{norm}$, Eqs. (\ref{ss_211}), (\ref{ss_21}). Expanding then to first order in $\dot \phi$ we obtain
\beq
{\cal S}_{norm} = {\cal S}_0  + \frac{i}{2} \int d \tau \int d{\bf r}  {\dot \phi} ({\bf r}, \tau)   Tr [G_0 ({\bf r}, \tau;{\bf r},\tau) \sigma_z]
\label{ss_3}
\eeq
where ${\cal S}_0$ does not depend on $\phi$. Introducing Fourier transformation for relative time and relative coordinate, replacing the integral  over momentum by $N_0 \int d \xi$, and keeping $e^{\pm i\omega \delta} $ factors (with infinitesimally small $\delta >0$) for particle and hole components of the Nambu Green's function in the normal state, we obtain for the second term in Eq. (\ref{ss_3})
\bea
\frac{i}{2} \int d \tau \int d{\bf r}  {\dot \phi} ({\bf r}, \tau)   Tr [G_0 ({\bf r}, \tau;{\bf r},\tau) \sigma_z] &=& \frac{i}{2}  \int d\tau \int d{\bf r}\, {\dot \phi} ({\bf r}, \tau) N_0\int_{-\mu}^\infty d\xi  \int \frac{d\omega}{2\pi} \left( \frac{e^{i\omega \delta}}{i \omega - \xi} -  \frac{e^{-i\omega \delta}}{i \omega + \xi}\right) \nonumber \\
&=&  i \int d\tau \int d{\bf r}\, {\dot \phi} ({\bf r}, \tau)  N_0  \int_{-\mu}^\infty d\xi \theta (-\xi)  =  i \frac{n_0}{2} \int d\tau \int d{\bf r}\, {\dot \phi} ({\bf r}, \tau)\,.
\label{chn_5}
\eea
Then
\beq
{\cal S}_{norm} = {\cal S}_0  + i \frac{n_0}{2} \int d\tau \int d{\bf r}\, {\dot \phi} ({\bf r}, \tau)\,.
\label{ss_3_a}
\eeq
Combining ${\cal S}_{1}$  from (\ref{b2}) and ${\cal S}_{norm}$, we obtain
\beq
{\cal S}_{1} + {\cal S}_{norm} = {\cal S}_0  + \frac{i}{2}   \int d\tau  n(\tau) \int d{\bf r}\, {\dot \phi} ({\bf r}, \tau)\,.
\label{ss_3_b}
\eeq

\subsubsection{The action ${\cal S}_2$: the ${\dot \phi}^2$ term}

\paragraph{Contribution from Eq. (\ref{S2}):\\}

To obtain the ${\dot \phi}^2$ term from Eq. (\ref{S2}) we we need the matrix elements $\langle\chi_{n,\lambda}|\partial_\tau\chi_{m,\lambda}\rangle$ between particle and hole states. Using Eqs.(\ref{chi0-}, \ref{chi0+}) we obtain after straightforward algebra
\begin{equation}\label{phi211}
{\cal S}_2 = \int d \tau \int d{\bf r}d{\bf r}^\prime \,\dot\phi({\bf r},\tau)\dot\phi({\bf r}^\prime,\tau) B({\bf r}-{\bf r}^\prime)\,,
\end{equation}
where
\begin{equation}\label{B}
B({\bf r}-{\bf r}^\prime) = {1\over S^2}\sum_{{\bf k},{\bf k}^\prime}\, \Big[{E_{\bf k}E_{{\bf k}^\prime}- \xi_{\bf k}\xi_{{\bf k}^\prime}+\Delta_0^2 \over  E_{\bf k}E_{{\bf k}^\prime}(E_{\bf k}+E_{{\bf k}^\prime})} - {|\xi_{\bf k}||\xi_{{\bf k}^\prime}|- \xi_{\bf k}\xi_{{\bf k}^\prime} \over  |\xi_{\bf k}| |\xi_{{\bf k}^\prime}|(|\xi_{\bf k}|+|\xi_{{\bf k}^\prime}|)} \Big]\, e^{i({\bf k}-{\bf k}^\prime)({\bf r}-{\bf r}^\prime)}\,,
\end{equation}
where the last term due to the $\Delta\rightarrow 0$ term in Eq. (\ref{S2}).
In the long wavelength limit $B({\bf r}-{\bf r}^\prime)$ can be approximated as $B_0\delta({\bf r}-{\bf r}^\prime)$, where
\begin{equation}\label{B10}
B_0 = {1\over S^2}\sum_{{\bf k},{\bf k}^\prime}\,\Big[{E_{\bf k}E_{{\bf k}^\prime}- \xi_{\bf k}\xi_{{\bf k}^\prime}+\Delta_0^2 \over  E_{\bf k}E_{{\bf k}^\prime}(E_{\bf k}+E_{{\bf k}^\prime})} - {sign(-\xi_{\bf k}) - sign(-\xi_{{\bf k}^\prime})\over \xi_{\bf k}-\xi_{{\bf k}^\prime}}\Big]\, (2\pi)^2\delta({\bf k}-{\bf k}^\prime)\,.
\end{equation}
In Eq. (\ref{B10}) we have rewritten the last term in the brackets of Eq. (\ref{B}), which corresponds to the familiar susceptibility of a normal (free) electron gas. This contribution is, however,  cancelled out  by the
second order contribution from ${\cal S}_{norm}$ in Eq. (\ref{ss_211}),
\begin{equation}\label{s_319}
{\cal S}_{norm}^{(2)} = {\cal S}_{norm}^{(1)} + (1/8)Tr[{\hat G}_0\sigma_z\dot\phi{\hat G}_0\sigma_z\dot\phi]\,,
\end{equation}
where we have expanded the logarithm up to the second order in $\dot\sigma$. The Green's function ${\hat G}_0$ in Eq. (\ref{s_319}) can be written in (Fourier representation) as
\begin{equation}\label{s_320}
{\hat G}_0(\omega, {\bf k}) =
\begin{bmatrix}
(i\omega - \xi_{\bf k})^{-1} & 0 \\
0 & (i\omega + \xi_{\bf k})^{-1}
\end{bmatrix}\,,
\end{equation}
where we have dropped $\nabla\phi$-dependent terms as they lead to higher (than the second) order contributions. Then we obtain that
\begin{equation}\label{s_321}
Tr[{\hat G}_0\sigma_z\dot\phi{\hat G}_0\sigma_z\dot\phi] = \int {d{\bf q}\over (2\pi)^2} \int {d\omega \over 2\pi} |\omega\phi(\omega, {\bf q})|^2\int {d\Omega \over 2\pi} \int {d{\bf k}\over (2\pi)^2}\Big[{1\over i\Omega-\xi_{\bf k}}{1\over i(\Omega+\omega)-\xi_{{\bf k}+{\bf q}}}+ {1\over i\Omega+\xi_{\bf k}}{1\over i(\Omega+\omega)+\xi_{{\bf k}+{\bf q}}} \Big]\,.
\end{equation}
Integrating over $\Omega$ and taking low frequency limit , i.e., setting $\omega=0$ in the resulting expression, one obtains
\bea\label{s_322}
Tr[{\hat G}_0\sigma_z\dot\phi{\hat G}_0\sigma_z\dot\phi] =\int {d{\bf q}\over (2\pi)^2} \int {d\omega \over 2\pi} |\omega\phi(\omega, {\bf q})|^2 \int {d{\bf k}\over (2\pi)^2} \Big[{sign(-\xi_{\bf k}) - sign(-\xi_{{\bf k}+{\bf q}})\over \xi_{\bf k}-\xi_{{\bf k}+{\bf q}}}\Big]\nonumber\\
= \int d\tau |\dot\phi({\bf q})|^2 \int {d{\bf q}\over (2\pi)^2} {d{\bf k}\over (2\pi)^2} \Big[{sign(-\xi_{\bf k}) - sign(-\xi_{{\bf k}+{\bf q}})\over \xi_{\bf k}-\xi_{{\bf k}+{\bf q}}}\Big] \,.
\eea
It is clear now that the second order term in the right hand side of Eq. (\ref{s_319}) is exactly the negative of the contribution produced by the last term in the brackets in Eqs. (\ref{B}, \ref{B10}) and therefore only the first term in the RHS of Eq. (\ref{B10}) contributes.

Performing integration over the momenta, we obtain
\begin{equation}\label{phi221}
{\cal S}_2 = \int d \tau N_0\Big(1+{\mu\over \sqrt{\mu^2+\Delta^2 (\tau)}}\Big)\,\int d {\bf r}\, {\dot\phi^2\over 8}\,.
\end{equation}

\paragraph{Another contribution to the prefactor for the $\dot\phi^2$ term:\\}

We now show that fluctuations of  $|\Delta (\tau)|$ give rise to an additional term in the action, $S_{extra}$, with
the same structure as in Eq. (\ref{phi221}).  For this we note that  the components of actions ${\cal S}_0$, ${\cal S}_1$, etc., are expressed in terms of fluctuating $|\Delta (\tau)|$ rather than in terms of constant $\Delta_0$. That is, the magnitude  of $\Delta$ fluctuates around its equilibrium value $\Delta_0$: $|\Delta(\tau)|=\Delta_0+\delta\Delta(\tau)$, and these longitudinal fluctuations are present in the action ${\cal S} = {\cal S}_0 + {\cal S}_1 + {\cal S}_2$.  They are small at weak coupling and are not important for the  $(\nabla\phi)^2$ in Eq. (\ref{totphifin}) and for the $\dot\phi^2$ term in Eq. (\ref{phi221}), but they give rise to $(\delta\Delta)^2$ term in ${\cal S}_0$, coming from ${\cal S}_{0,a} (\Delta)$ in Eq. (\ref{E0}) and to $\dot\phi\delta\Delta$ term in ${\cal S}_1$, coming from expanding the prefactor for $\dot \phi$ term in (\ref{b2}) in $\delta\Delta(\tau)$. The combination of these two pieces gives rise to the additional ${\dot\phi}^2$  term in the action, which we now compute.

Within our approximation, $\Delta(\tau)$ is independent of $\bf r$, hence one can simply expand Eqn. (\ref{E0}) to the second order in $\delta\Delta$. The   linear term is zero because $\Delta_0$ corresponds to the minimum in the free energy, but the second order term is finite. Using Eqs. (\ref{self}, \ref{Delta0}, \ref{Delta1}), we obtain after some straightforward algebra  that
\begin{equation}\label{DeltaE1}
{\cal S}_{0,a} (\Delta) = {\cal S}_{0,a} (\Delta_0) + S N_0 \int d\tau {\Delta_0^2\over 4E_0(E_0+E_F)} (\delta\Delta)^2 \,,
\end{equation}
where, we recall,  $S$ is the area of a 2d sample.

Similarly,  we expand  in Eq. (\ref{b2})  to  linear order in $\delta\Delta(\tau)$, use Eqs. (\ref{Delta0}) and (\ref{Delta1}), and  obtain ${\cal S}_1$ in terms of $\Delta_0$ with the extra term with the product of first derivatives:
\begin{equation}\label{berry11}
{\cal S}_1= - (iN_0/2)\,\int d\tau \int d{\bf r}\,\Big[(\sqrt{\mu^2+\Delta_0^2}-|\mu|)\dot\phi + {\Delta_0\over E_0+E_F}\dot\phi\delta\Delta\Big]\,.
\end{equation}

Combining the last terms in (\ref{DeltaE1}) and (\ref{berry11}) together, we obtain the extra piece in the action, $\delta {\cal S}$,  associated with longitudinal gap fluctuations:
\begin{equation}\label{wed_1}
\delta {\cal S} = {N_0\over 2}\,\int d\tau \int d{\bf r} \Big[{\Delta_0^2\over 2E_0(E_0+E_F)}(\delta\Delta)^2 - i{\Delta_0\over E_0+E_F}\dot\phi\delta\Delta\Big]\,.
\end{equation}
Averaging over the Gaussian fluctuations of $\delta\Delta$ (which is the same as completing the square in (\ref{wed_1}))  we obtain
an additional contribution to  the action, ${\cal S}_{extra}$,  in the form
\begin{equation}\label{phi22}
{\cal S}_{extra} =  \int d \tau N_0\Big(1-{\mu\over \sqrt{\mu^2+|\Delta_0|^2}}\Big)\,\int d {\bf r} {\dot\phi^2\over 8}\,,
\end{equation}
Combining this with $ {\dot\phi}^2$ term in ${\cal S}_2$ in (\ref{phi221}) we obtain
\begin{equation}\label{phi23}
{\cal S}_2 + {\cal S}_{extra}  = N_0 \int d \tau \int d{\bf r}\,{\dot\phi^2\over 4}\,.
\end{equation}

\subsection{The full long-wavelength action}

Combining the $(\nabla \phi)^2$ and $\dot \phi^2$ terms,  Eqs. (\ref{totphifin}) and (\ref{phi23}), we obtain the regular part of the action in the form
\begin{equation}
\label{wed_2}
{\cal S}_{reg} =  N_0 \int d \tau \int d{\bf r}\, \Big[\big(\big(\sqrt{\mu^2+|\Delta (\tau)|^2} + \mu\big) \big) \frac{(\nabla\phi)^2}{8m}  + {\dot\phi^2\over 4} \Big]\,.
\end{equation}
To our accuracy, the prefactor for $(\nabla\phi)^2$ term can be evaluated at $|\Delta (\tau)|^2= \Delta^2_0$. We then obtain
\begin{equation}
\label{wed_3}
{\cal S}_{reg} =   N_0 \int d \tau \int d{\bf r}\, \Big[{E_F\over 4m}(\nabla\phi)^2 + {\dot\phi^2\over 4} \Big]\,.
\end{equation}
The coefficient in front of $(\nabla\phi)^2$ in Eq. (\ref{wed_3}) can be rewritten as more familiar $n/8m$ (Ref. \cite{rho}). Transforming to Fourier components (momentum $q$ and  real frequency $\Omega$), we obtain from (\ref{wed_3})  ${\cal S}_{cont} \propto \sum_{q,\Omega} |\phi_{q,\Omega}|^2 (\Omega^2 - q^2 v^2_F/2 )$.  The prefactor for $ |\phi_{q,\Omega}|^2$ is the inverse susceptibility of phase fluctuations. We see that it has a pole at $\Omega = \pm (v_F/\sqrt{2}) q$. The pole position corresponds to the frequency of a gapless phase fluctuation mode, whose velocity is $v_F/\sqrt{2}$, independent on the ratio of $E_F/E_0$.

The full term  linear in ${\dot \phi}$ (the Berry phase term) is obtained by setting $\Delta (\tau) = \Delta_0$ in Eq. (\ref{ss_3_b}):
\beq
{\cal S}_{Berry} = i A \int d\tau \int d{\bf r} {\dot \phi}
\label{wed_4}
\eeq
where
\beq
A =   \frac{n}{2} = \frac{N_0}{2}~\left(\sqrt{\mu^2+\Delta_0^2} + \mu\right)\,
\label{wed_5}
\eeq
and $n$ is the actual electron density.

The result for $A$ agrees with  Refs. \cite{thouless,jacek,gaitan,simanek_2,kopninvolovik}. Note, however, that there is one element in our calculation, which has not been emphasized in earlier works. Namely, the absence of $n_0$ in (\ref{wed_5}) could be interpreted as if there is no contribution from $\Delta \to 0$. We argue that this is not entirely true. In our calculation, there are two contributions from $\Delta \to 0$:  the term ${\cal S}_1$ in the limit $\lambda =0$ and the term ${\cal S}_{norm}$. The  $n_0/2$ pieces from these two terms do cancel out,  however, the full contribution to $A$ from $\Delta \to 0$ does not vanish and gives $(1/2) \sum_k \left[\xi_k/|\xi_k| + (1-\xi_k/|\xi_k|)\right] = (1/2)\sum_k (1)$. This formally divergent piece cancels out the divergence in the contribution to the prefactor from  ${\cal S}_1$ at $\lambda =1$ (i.e., at non-zero $\Delta$), which is $(-1/2) \sum_k (\xi_k/E_k) = n/2 - (1/2) \sum_k (1)$.  Without this cancellation, the coefficient for ${\dot \phi}$ term would contain a parasitic, formally infinite term.  The same  holds for the coefficient for the $(\nabla \phi)^2$ term in the action: if we were to neglect the contributions from $\Delta \to 0$,  the prefactor would be $(-1/8m) \sum_k (\xi_k/E_k) = (1/8m) (n - \sum_k (1))$. The parasitic $\sum_k (1)$ term is canceled out by the sum of the two contributions from $\Delta \to 0$, as we showed above.

If $\phi$ is well defined for all ${\bf r}$ and $\tau$, the Berry phase term reduces to the contribution from the boundary and does not affect the dynamics. The situation changes when $\phi$ is  singular, as in the case of a moving vortex with coordinates $X(\tau)$ and $Y(\tau)$.   Then the Berry phase term in the action becomes proportional to $\int d \tau X(\tau) {\dot Y} (\tau) - Y(\tau) {\dot X} (\tau)$, which cannot be expressed as a total derivative and contributes to the vortex dynamics.  We show that the action (\ref{wed_4}) describes the contribution to the Berry term from fermions far away from the vortex core. We show that there is another contribution, which comes from the states right at the center of the vortex core. This last term originates from  $\nabla \phi$ terms in $S_{norm}$.

\section{The Berry phase term in the action for a moving vortex}

The order parameter for a moving vortex in 2d can be written as
\begin{equation}\label{vort}
\Delta (\tau, {\bf r}) = \Delta [{\bf r}-{\bf R}(\tau)] = |\Delta [{\bf r}-{\bf R}(\tau)]| e^{i\phi[{\bf r}-{\bf R}(\tau)]}\,,
\end{equation}
where ${\bf R}(\tau)$ is vortex center,
\begin{equation}\label{vortphase}
\phi(\tau,{\bf r})={\rm tan}^{-1}\Big[{y-Y(\tau)\over x-X(\tau)}\Big]\,,
\end{equation}
and $|\Delta({\bf r})|\rightarrow \Delta_0$ for $r\gg \lambda$, where $\lambda$ is the penetration depth.

The spectrum of the Bogolubov-DeGennes equation (Eq. (\ref{bdgmain})) near a vortex has been extensively studied~\cite{caroli,bardeen,simanek,jacek,simanek_2,stone,gaitan,ao_1} and is known to posses both continuous and discrete branches corresponding to delocalized and localized eigenstates, respectively.  The localized eigenstates are known as  Caroli, de Gennes, Matricon states~\cite{caroli}. The continuous part of the spectrum covers the range $|E_n^c|>\Delta_0$, while discrete states have energies $|E_n^d|<\Delta_0$.

The contributions to the vortex motion come from the occupied states with negative energies. A  generic eigenstate, corresponding to
$E_n <0$, can be expressed as
\begin{equation}\label{chibound}
|\chi^{-}_v ({\bf r})\rangle =  e^{i ({\hat\sigma}_z/2)\phi({\bf r}-{\bf R})} |{\tilde \chi}^{-}_\nu({\bf r})\rangle; ~~|{\tilde \chi}^{-}_\nu({\bf r})\rangle =  e^{i \nu\phi({\bf r}-{\bf R})}
\begin{bmatrix}
{\tilde v}_\nu(|{\bf r}-{\bf R}|) \\
-{\tilde u}_\nu(|{\bf r}-{\bf R}|)
\end{bmatrix}
\ ,
\end{equation}
where
$\nu = n+1/2$ and $n$ is an integer, $\phi({\bf r}-{\bf R})$ is given by Eq. (\ref{vortphase}),  and the radial functions $u_\nu^{b}(r)$ and $v_\nu^{b}(r)$ {\it and their derivatives} with respect to $r$ are continuous for all $r$.

The eigenfunctions for the localized states are proportional to  $J_{|\nu\pm1/2|}(k_F\,r)$ at small $r\ll\lambda$ (upper sign for ${\tilde u}_\nu$, lower for ${\tilde v}_\nu$).  At large distances, when $r\gg\lambda$, both $|{\tilde v}_\nu|^2$ and $|{\tilde u}_\nu|^2$ decay exponentially (Ref. \cite{simanek}). The eigenfunctions for continuous states  are expressed via $J_{|\nu\pm1/2|}(k\,r)$ at small $r\ll\lambda$ (where $k$ is generally a function of $\nu$), while for $r\gg\lambda$  they are parameterized by $\nu$ and momentum $k$, which becomes an independent variable (Refs. \cite{caroli,bardeen,ao_1}):
\bea
{\tilde u}_{\nu} (r) = {\tilde u}_{\nu,k} (r) =
u_k J_{|\nu|}(k r),~
{\tilde v}_{\nu} (r) = {\tilde v}_{\nu,k} (r)  = v_k J_{|\nu|}(k r)
\label{aa_3}
\eea
where
\begin{equation}
u_k = \left(\frac{1}{2} + \frac{\xi_k}{2|E^{(-)}_k|}\right)^{1/2},~  v_k = \left(\frac{1}{2} - \frac{\xi_k}{2|E^{(-)}_k|}\right)^{1/2},
\end{equation}
and $E^{(-)}_k = -\sqrt{\xi^2_k + |\Delta_0|^2}$.
The full solution of the Bogolubov-DeGennes equation  for a vortex is expressed via Hankel functions, which are linear combinations of Bessel functions and Neumann functions. The Neumamm functions $Y_{|\nu\pm 1/2|} (x)$ and $Y_{|\nu|} (x)$, however, grow when the index $\nu$ becomes larger than the argument $x$, and the sums over $\nu$ in (\ref{chna_9}) do not converge. The Bessel functions $J_{|\nu|} (x)$, on the contrary, decay exponentially when $\nu$ gets larger than $x$.  By this reason, we only consider the solutions expressed via the Bessel functions.

Because both localized and extended states are present, specified by a discrete parameter $\nu$,  it is not a'priori guaranteed that we can use the results from the previous section, which were obtained using the eigenfunctions far away from the vortex core, when $\nu$ can be treated as a continuous variable.

In this section we re-evaluate the prefactor for ${\dot \phi}$ term  using the exact eigenfunctions $|\chi^{-}_\nu ({\bf r})\rangle$.  We first re-evaluate the terms ${\cal S}_1$ and ${\cal S}_{norm}$ and show that they are determined by fermions far away from the vortex core and have the same forms as we found in the previous section. Then we take a closer look at seemingly innocent part of $S_{norm}$, which does not contain ${\dot \phi}$, but does depend on ${\nabla \phi}$. We argue that it  also contributes to the Berry phase term for a moving vortex, and this contribution comes from fermions inside the vortex core.

\subsection{The ${\cal S}_1$ term for the vortex motion}

The  ${\cal S}_1$ term in the action is given by Eq. (\ref{S1}), which is valid for arbitrary $|\chi^{-}_\nu ({\bf r})\rangle$.
To obtain ${\cal S}_1$  for a vortex  we  need to evaluate
\beq
\sum_\nu \int d{\bf r} \langle\chi^{-}_{\nu}(\tau,{\bf r})|\partial_\tau\chi^{-}_{\nu}(\tau, {\bf r})\rangle\,.
\label{chna}
\eeq
with $|\chi^{-}_\nu ({\bf r})\rangle$ from (\ref{chibound}). Substituting these  $|\chi^{-}_\nu ({\bf r})\rangle$  into (\ref{chna}) we obtain
\beq
\sum_\nu \int d{\bf r} \langle\chi^{-}_{\nu}(\tau,{\bf r})|\partial_\tau\chi^{-}_{\nu}(\tau, {\bf r})\rangle =  i \int d{\bf r}\, \Phi ({\bf r}-{\bf R}(\tau)) \partial_{\tau} \phi[{\bf r}-{\bf R}(\tau)]\,,
\label{chna_1}
\eeq
where
\beq
\Phi (|{\bf r}|) =\sum_\nu \left[|{\tilde u}_\nu ({\bf r})|^2
\left(\nu -\frac{1}{2}\right) + |{\tilde v}_\nu ({\bf r})|^2 \left(\nu +\frac{1}{2} \right)\right].
\label{chna_9}
\eeq
Using (\ref{vortphase}), one can re-express $\partial_{\tau} \phi[{\bf r}-{\bf R}(\tau)]$ as
\beq
\partial_{\tau} \phi[{\bf r}-{\bf R}(\tau)] = {\dot X (\tau)} \left[\frac{y-Y(\tau)}{(x-X(\tau))^2 + (y-Y(\tau))^2}\right] -{\dot Y (\tau)} \left[\frac{x-X(\tau)}{(x-X(\tau))^2 + (y-Y(\tau))^2}\right]\,,
\label{chna_2}
\eeq
such that
\beq
\sum_\nu \int d{\bf r} \langle\chi^{-}_{\nu}(\tau,{\bf r})|\partial_\tau\chi^{-}_{\nu}(\tau, {\bf r})\rangle = i \left[{\dot X (\tau)} Q_x - {\dot Y (\tau)} Q_y\right]\,,
\label{chna_3}
\eeq
where
\bea
Q_x &=& \int d {\bf r} \left[\frac{y-Y(\tau)}{(x-X(\tau))^2 + (y-Y(\tau))^2}\right] \Phi ({\bf r}-{\bf R}(\tau)) \,,\nonumber \\
Q_y &=& \int d {\bf r} \left[\frac{x-X(\tau)}{(x-X(\tau))^2 + (y-Y(\tau))^2}\right] \Phi ({\bf r}-{\bf R}(\tau))\,.
\label{chna_4}
\eea

We show below that $Q_x \propto Y (\tau)$ and $Q_y \propto X (\tau)$. It is then tempting to compute the prefactors by evaluating the derivatives $d Q_y/d X$ and $d Q_x/dY$. This has to be done with extra care as the integrals for  $d Q_y/d X$ and $d Q_x/dY$ are infra-red singular and have to be properly regularized (see below).  We use a different computational procedure and evaluate the integrals in (\ref{chna_4}) directly assuming that our system has a finite size $2D$ in both $x$ and $y$ directions. We show that $Q_x$ and $Q_y$   remain finite if we set $D$ to infinity at the end of the calculation. We verified that the result does not depend on the  We checked that the result does not depend on the geometry of the integration range, as long as the symmetry between $x$ and $y$ is preserved, i.e., $Q_x$ and $Q_y$  remains the same if we assume that the boundary of our system is, e.g., a circle instead of a square.

Let's  evaluate $Q_x$ first.  Shifting the variables of integration from $x$ and $y$ to ${\tilde x} = x - X(\tau)$ and ${\tilde y} = y - Y(\tau)$, we obtain from (\ref{chna_4})
\beq
Q_x = \int_{-D-X(\tau)}^{D-X(\tau)} d{\tilde x} \int_{-D-Y(\tau)}^{D-Y(\tau)} d{\tilde y} \left[\frac{{\tilde y}}{{\tilde x}^2 + {\tilde y}^2}\right] \Phi ({\tilde r})\,,
\label{chna_5}
\eeq
where ${\tilde r} = ({\tilde x}^2 + {\tilde y}^2)^{1/2}$. Using that the integrand is odd in ${\tilde y}$, we re-write (\ref{chna_5}) as
\beq
Q_x = -\int_{-D-X(\tau)}^{D-X(\tau)} d{\tilde x} \int_{D-Y(\tau)}^{D+Y(\tau)} d{\tilde y} \left[\frac{{\tilde y}}{{\tilde x}^2 + {\tilde y}^2}\right] \Phi ({\tilde r})\,.
\label{chna_6}
\eeq
This shows that the result comes from a tiny range of ${\tilde y}$ around ${\tilde y} =D$.  A simple experimentation then shows that typical ${\tilde x}$ are also of order $D$.
Assuming that $\Phi ({\tilde r})$ tends to the value at ${\tilde r} \sim D$, independent on the ratio ${\tilde x}/{\tilde y}$, we pull $\Phi (D)$ from the r.h.s. of (\ref{chna_6}).  The integration over ${\tilde x}$ is then elementary, and the result is
\beq
Q_x = - \pi Y (\tau) \Phi (D)\,.
\label{chna_7}
\eeq
Evaluating  $Q_y$ the same way, we find $Q_y= - \pi X(\tau) \Phi (D)$. Substituting $Q_x$ and $Q_y$ into (\ref{chna_3}), we obtain
\beq
\sum_\nu \int d{\bf r} \langle\chi^{-}_{\nu}(\tau,{\bf r})|\partial_\tau\chi^{-}_{\nu}(\tau, {\bf r})\rangle = i\pi \Phi (D)
\left[X(\tau) {\dot Y (\tau)}  -  Y(\tau) {\dot X (\tau)} \right]\,.
\label{chna_8}
\eeq

We emphasize that $Q_x$ and $Q_y$ are determined by distances of order $D$, i.e., the contribution comes from fermions far away from the vortex core.  There is no contribution from $r=0$,  contrary to what has been reported in some earlier papers (see e.g. Ref. \cite{jacek}).  In these earlier works the authors computed $d Q_y/d X$ and $d Q_x/dY$ by differentiating only in the term in the brackets  (\ref{chna_4}) (i.e., not differentiation $\Phi ({\bf r} - {\bf R} (\tau))$), and set $X = Y=0$ in the integrand  before evaluating the integral over $d{\bf r}$. Then the result comes from the smallest ${\bf r} =0$, as we will see below. However, the full $dQ_x/dY$ contains also the derivative of $\Phi ({\bf r} - {\bf R} (\tau))$), i.e., if we differentiate under the integral in (\ref{chna_4}) and set $X=Y=0$, we obtain
\bea
\frac{dQ_x}{dY} = \int_{-D}^D dx dy \Big(\left[-\frac{1}{x^2 +y^2} + 2 \frac{y^2}{(x^2+y^2)^2}\right] \Phi (x,y) - \frac{y}{x^2+y^2} \frac{\partial \Phi (x,y)}{\partial y}\Big)
\label{aa_1}
\eea
The first term is formally zero (it contains $y^2-x^2$ as the overall factor), but it also diverges at $x=y=0$. To regularize this term, we introduce an infinitesimally small "mass" term in the denominator, i.e., replace $x^2 +y^2$ by $x^2 + y^2 + \epsilon^2$, evaluate the integral with a finite $\epsilon$, and then set it to zero. Transforming to polar coordinates $x =r \cos{\theta}$, $y = r \sin{\theta}$ and using $\partial \Phi (x,y)/\partial y = (y/r) d \Phi/dr = sin{\theta} d \Phi/dr$, we then obtain from (\ref{aa_1})
\bea
\frac{dQ_x}{dY} =   -2\pi \epsilon^2 \int \frac{r dr}{(r^2 + \epsilon^2)^2} \Phi (r) -\pi \int_0^D dr \frac{d \Phi (r)}{dr}
\label{aa_2}
\eea
The evaluation of the integral is now elementary. In the first term the integral comes from $r \sim \epsilon$ and cancels $\epsilon^2$ in the numerator ($r \epsilon^2/(r^2+ \epsilon^2)^2$ acts as $\delta (r)$). The term then yields $-\pi \Phi (0)$. This is what has been obtained in Ref.  \cite{jacek}) and earlier papers cited in that work. The full result, however, also contains the contribution from the second term. It obviously gives  $-\pi (\Phi (D) - \Phi (0))$.  The sum of the two terms  is $-\pi \Phi (D)$, with no contribution from ${\bf r} =0$. This  agrees with (\ref{chna_7}).

Substituting Eq. (\ref{chna_8}) into Eq. (\ref{S1}), we find
\begin{equation}\label{Bvortfin}
{\cal S}_1 = i \pi A_{1}
\int d \tau  \left( X(\tau) {\dot Y (\tau)}  -  Y(\tau) {\dot X (\tau)} \right)\,.
\end{equation}
where $A_{1} =\Phi^\Delta (D) -\Phi^{\Delta \to 0} (D)$.
We emphasize that $ X(\tau) {\dot Y (\tau)}  -  Y(\tau) {\dot X (\tau)}$  is {\it not} a full derivative, hence ${\cal S}_1$ term does contribute to the equation  of motion for a vortex.

To obtain $\Phi^\Delta (D) \equiv \Phi (r=D)$ we need to know the forms of ${\tilde u}_\nu$ and ${\tilde v}_\nu$ at large $r$. The eigenfunctions for localized  eigenstates decay exponentially when $r \sim D$ and hence are irrelevant for our purpose. The eigenfunctions for continuous states with a negative energy at distances larger than the penetration depth are given by Eq. (\ref{chi0-}).

Using these forms,  we obtain
\bea
&&
\Phi (x) = \sum_{\nu,k} \left[u^2_k (\nu-
1/2) J^2_{|\nu|} (x)  + v^2_k (\nu+
1/2) J^2_{|\nu|} (x)\right] \nonumber \\
&=&  \sum_k \left( \left[(u^2_k +v^2_k) \sum_{\nu} (\nu-
1/2) J^2_{\nu} (x)\right]+v^2_k  \sum_{\nu} J^2_{|\nu|} (x)\right)
\eea
To evaluate the $\sum_{\nu} J^2_{|\nu|} (x)$, we note  that at large $x$ and $\nu <x$, the Bessel function can be approximated as
\beq
J_{\nu} (x) \approx \sqrt{\frac{2}{\pi}}  \frac{1}{(x^2-\nu^2)^{1/4}} \sin{\left(x + \frac{\nu^2}{2x} - \frac{\pi}{4} (2\nu+1)\right)}
\label{aa_4}
\eeq
This formula is valid up to $\nu = x - O(x^{1/3})$.  The sum over $\nu$ is determined by large $\nu = O(x)$, for which the summation over $\nu$ can be replaced by integration.  We assume and then verify that the integral is determined by $\nu = O(x)$, but $x-\nu \gg x^{1/3}$.  The contribution from this range is
\beq
2 \int_0^{x - O(x^{1/3})}  J^2_\nu (x) = \frac{2}{\pi} \int_0^{x - O(x^{1/3})} \frac{dx}{\sqrt{x^2-\nu^2}} = 1 - O(x^{-1/6})\,.
\label{aa5}
\eeq
One can easily verify that the contribution from $|\nu-x| \leq x^{1/3}$  scales as $x^{-1/6}$, and the contribution from larger $\nu> x + O(x^{1/3})$ is even smaller because at such $\nu$, $J_{|\nu|} (x)$ decays exponentially. Then  $\sum_{\nu} J^2_{|\nu|} (x) =1$  up to corrections, which vanish at $x \to \infty$

Further, $\sum_{\nu} (\nu J^2_{|\nu|} (x))$ vanishes because of cancellation between terms with terms with positive and negative $\nu$. As the consequence,
\beq
\left[(u^2_k +v^2_k) \sum_{\nu} (\nu
- 1/2) J^2_{\nu} (x)\right]+v^2_k  \sum_{\nu} J^2_{|\nu|} (x) =
-\frac{1}{2}+v^2_k
\eeq
and, hence,
\beq
\Phi^\Delta (D) =-\frac{1}{2} \sum_k (1)
+\sum_k \left(v^2_k\right)^\Delta\,.
\label{aa6}
\eeq
This formula could also be obtained if we assumed from the beginning that  at large ${\bf r}$, the eigenfunctions for the continuous spectrum approach those for a superconductor with a constant gap $\Delta$, i.e., radial quantum number $\nu$ becomes momentum $k$, and ${\tilde \chi}^{-}_{\nu}(\tau,{\bf r}) \rangle$ becomes
\begin{equation}\label{chi0-1}
|{\tilde\chi}_{\bf k}^{(-)}({\bf r})\rangle =
\begin{bmatrix}
v_k
 \\
-u_k
\end{bmatrix}
{e^{i{\bf kr}}\over \sqrt{S}} .
\end{equation}
The second term in (\ref{aa6}) can be easily evaluated
\beq
\sum_k \left(v^2_{\bf k}\right)^\Delta = \frac{N_0}{2} \int_{-\mu}^\infty d \xi\,\left(1 - \frac{\xi}{\sqrt{\xi^2 + \Delta^2_0}}\right) = \frac{N_0}{2} \Big(\sqrt{\mu^2+\Delta_0^2}+\mu\Big) = \frac{n}{2}\,,
\label{chna_10_1}
\eeq
where, we remind, $n = 2N_0 E_F$ is the actual density of fermions.  However, the first  term in (\ref{aa6}) is the sum over all momenta, and is formally infinite.
We now recall that the prefactor in the $S_1$ term in the action, Eq. (\ref{Bvortfin}), contains the difference
$  \Phi^\Delta (D) - \Phi^{\Delta \to 0} (D)$. We assume that the distances $r \sim D$ are  outside the vortex core even when $\Delta \to 0$. Then $\Phi^{\Delta \to 0} (D)$ is determined by the same Eq. (\ref{aa6}) as $\Phi^\Delta (D)$, the only difference is that now
\begin{equation}
u_k =  \left(\frac{1}{2} + \frac{\xi_k}{2|\xi_k|}\right)^{1/2},~  v_k = \left(\frac{1}{2}  - \frac{\xi_k}{2|\xi_k|}\right)^{1/2},
\end{equation}
and
\beq
\sum_k \left(v^2_{\bf k}\right)^{\Delta \to 0} = \frac{N_0}{2} \int_{-\mu}^\infty d \xi \left(1 - \frac{\xi}{|\xi|} \right) = \frac{N_0}{2} \Big(|\mu|+\mu\Big) = \frac{n_0}{2}\,,
\label{chna_11}
\eeq
where $n_0$ is the density of free fermions at the same chemical potential $\mu$.  Accordingly,
\beq
\Phi^{\Delta \to 0} (D) =-\frac{1}{2} \sum_k (1)
+  \sum_k \left(v^2_k\right)^{\Delta =0}\,.
\label{aa6_1}
\eeq
Combining (\ref{aa6}) and (\ref{aa6_1}), we obtain
 \beq
 A_{1} =  \Phi^\Delta (D) - \Phi^{\Delta \to 0} (D) = \sum_k \left(v^2_k\right)^{\Delta} - \sum_k \left(v^2_k\right)^{\Delta =0} = \frac{n-n_0}{2}\,.
 \label{chna_10}
 \eeq
We emphasize that this result is free from ultra-violet divergencies -- the term $\sum_k (1)$ cancels out between (\ref{aa6}) and (\ref{aa6_1}). To obtain this cancellation one has to keep the contributions to $A_1$ from  the action at $\Delta \to 0$.

\subsection{The ${\cal S}_{norm}$ term for the vortex motion}

Now we need to add to this result the contribution from ${\cal S}_{norm}$ (the term in the action at $\Delta \to 0$, proportional to ${\dot \phi}$). This contribution is computed in the same way as the one from ${\cal S}_1$. Namely, the contribution to $\partial \tau \phi[{\bf r} - {\bf R}]$ comes from fermions far away from the vortex core, hence we can just use ${\cal S}_{norm}$ from Eq. (\ref{chn_5}):
${\cal S}_{norm} = (i n_0/2) \int d \tau \int d{\bf r} \dot \phi ({\bf r}, \tau)$ and substitute
$\int d {\bf r} \dot \phi ({\bf r}, \tau) = \pi (\left( X(\tau) {\dot Y} (\tau) - Y(\tau) {\dot X} (\tau)\right)$.
This gives
\begin{equation}\label{Bvortfin_1}
{\cal S}_{norm} = i \pi A_{norm}\int d \tau  \left( X(\tau) {\dot Y (\tau)}  -  Y(\tau) {\dot X (\tau)} \right)\,,
\end{equation}
where $A_{norm} = n_0/2$.
Adding  $A_{norm}$ to  $A_1$ from (\ref{chna_10}),  we obtain at this stage the Berry phase term in the action in the form
\beq
{\cal S}_{Berry,1} = i \pi A_{vort,1} \int d \tau  \left( X(\tau) {\dot Y (\tau)}  -  Y(\tau) {\dot X (\tau)} \right)\,,
\label{aa8}
\eeq
where
\beq
A_{vort,1} = \frac{n}{2}\,.
\eeq
We see that the prefactor is the same as in the action in Eq. (\ref{wed_4}). This is not surprising because the contribution comes solely from the states well outside the vortex core, where the eigenfunctions can be approximated by  the  ones in the absence of a vortex.

\subsection{Contribution from the vortex core}

We now show that there is another contribution to the  Berry phase term in the action for a moving vortex, which comes from the vortex core.  This extra contribution  is "hidden" in the zero-order term in $S_{norm} = Tr \log(G_0^{-1})$ -- the one which does not contain ${\dot \phi}$.
We label this term as $S^0_{norm}$.  This term does, however, depend on $\nabla\phi$, because we remind that to eliminate $\Delta$ from the Green's function we had to  apply the unitary transformation $\hat U$ under $Tr \log$,  with  $\hat U$ given by (\ref{ss_1}).  Under this unitary transformation the kinetic energy operator $\hat\xi(\nabla) = -\nabla^2/(2m) $ changes to $\hat\xi(\nabla+ (i/2)\sigma_z\nabla\phi)$, where $\phi=\phi(\bf r-{\bf R} (\tau))$. As the result,  $S^0_{norm}$ does actually depend on ${\bf R}(\tau)$ via $\nabla\phi$.

Let's assume that ${\bf R} (\tau)$ is small and expand $S^0_{norm} ({\bf R}) = -Tr \log G_0^{-1}$  to second order  in  ${\bf R} (\tau)$.  A generic expansion yields
\begin{equation}\label{react1}
S^0_{norm} ({\bf R}) = S^0_{norm} ({\bf R}=0)  - \int d\tau \eta_{\alpha}(\tau)\, R_{\alpha} (\tau) - \int d\tau d\tau^\prime \eta_{\alpha\beta}(\tau-\tau^\prime)\,  R_{\alpha} (\tau) R_{\beta} (\tau^\prime) + ... \,.
\end{equation}
where the summation over repeated indices is assumed.

The first order response function $\eta_\alpha$ is zero because of translational invariance. To see this, we note that  the matrix $G_0$ is diagonal, and so $Tr \log G_0^{-1} = Tr \log (G_0^+)^{-1} + Tr \log (G_0^-)^{-1}$, where
\begin{equation}\label{react20}
(G_0^\pm)^{-1} = -\partial_\tau \pm \hat\xi(\nabla\pm (i/2)\nabla\phi)\,.
\end{equation}
Then
\begin{equation}\label{react2_1}
\eta_\alpha = Tr \big[G_0^+ {\partial{\hat\xi}^+\over \partial R_\alpha}\big] + Tr \big[G_0^- {\partial{\hat\xi}^-\over \partial R_\alpha}\big]
\end{equation}
where $G_0 = G_0 (r, \tau; r,\tau)$. In Fourier representation
\beq \label{react2}
\eta_\alpha
= \sum_n \int {d\omega\over 2\pi} \Big[ {\langle\chi_n^+|\partial{\hat\xi}^+/\partial R_\alpha|\chi_n^+\rangle\, e^{i\omega\delta}  \over i\omega - \xi_n^+} + {\langle\chi_n^-|\partial{\hat\xi}^-/\partial R_\alpha|\chi_n^-\rangle\, e^{-i\omega\delta} \over i\omega + \xi_n^-}\Big] \,,
\end{equation}
where $|\chi_n^\pm\rangle$ ($\xi_n^\pm$) are the eigenfunctions (eigenvalues) of $\hat\xi^{\pm}=\hat\xi(\nabla\pm (i/2)\nabla\phi)$ (we present explicit expressions below). The eigenvalues of $\xi_n^\pm$ do not dependent on $R_\alpha$ in a translationally invariant system, hence $\partial \xi_n^\pm /\partial R_\alpha = 0$. Accordingly,
$\langle\chi_n^\pm|\partial {\hat\xi}^\pm/\partial R_\alpha|\chi_n^\pm\rangle = \partial \langle\chi_n^\pm|{\hat\xi}^\pm|\chi_n^\pm\rangle/\partial R_\alpha = \partial \xi_n^\pm /\partial R_\alpha$ also vanish. Hence, $\eta_\alpha =0$.

The second order term in Eq. (\ref{react1})  is non-zero, as we will see. In the Fourier representation,
\beq
\int d\tau d\tau^\prime \eta_{\alpha\beta}(\tau-\tau^\prime)\,  R_{\alpha} (\tau) R_{\beta} (\tau^\prime) = \int \frac{d \omega}{2\pi} R_{\alpha} (-\omega) R_{\beta} (\omega) \eta_{\alpha \beta} (\omega)
\label{exx_1}
\eeq
The form of the Berry phase term is reproduced if we set $\eta_{\alpha \beta} (\omega) = \omega \epsilon_{\alpha \beta} {\bar \eta}$, where $\epsilon_{\alpha \beta}$ is antisymmetric tensor ($\epsilon_{xy} = - \epsilon_{yx} =1$).  Then
\beq
S^0_{norm} ({\bf R}) = S^0_{norm} ({\bf R}=0) - i \frac{\bar \eta}{2} \left(X(\tau) {\dot Y} (\tau) - Y(\tau) {\dot X} (\tau)\right)\,.
\label{exx_2}
\eeq
Our goal therefore is to extract linear in $\omega$ and antisymmetric in $\alpha, \beta$ contribution to $\eta_{\alpha \beta}$.  In $\tau$ space, $\eta_{\alpha \beta} = \omega \epsilon_{\alpha \beta} {\bar \eta}$ corresponds to $\eta_{\alpha \beta} (t) = i  \epsilon_{\alpha \beta} {\bar \eta} d \delta (t)/dt$ ($t = \tau - \tau'$).

There are two contributions to $\eta_{\alpha \beta} (\tau-\tau')$. One comes from the second order term in Taylor expansion of $\hat\xi({\bf R})$ and has the form
\begin{equation}\label{react3}
\eta_{\alpha\beta}^{(1)} (\tau - \tau') =  Tr \big[G_0 {\partial^2{\hat\xi}\over \partial R_\alpha\partial R_\beta}\big]\,\delta(\tau-\tau^\prime)\,,
\end{equation}
where again $G_0 = G_0 (r, \tau, r, \tau)$. In Fourier representation, $\eta_{\alpha\beta}^{(1)} (\omega) $ does not depend on $\omega$  ($\eta_{\alpha\beta}^{(1)}  (\omega) = \eta_{\alpha\beta}^{(1)} $) and is given by
\begin{equation}\label{react4}
\eta_{\alpha\beta}^{(1)} =  Tr \big[G_0 {\partial^2{\hat\xi}\over \partial R_\alpha\partial R_\beta}\big] = \sum_n \int \frac{d \omega_1}{2\pi}
\left[ \frac{\langle\chi_n^+|\partial^2{\hat\xi}^+/\partial R_\alpha\partial R_\beta|\chi_n^+\rangle}{i\omega_1 - \xi^+_n}  + \sum_n \int \frac{d \omega_1}{2\pi} \frac{\langle\chi_n^-|\partial^2{\hat\xi}^-/\partial R_\alpha\partial R_\beta|\chi_n^-\rangle}{i\omega_1 + \xi^-_n}\right] \,.
\end{equation}
Evaluating the integral over $\omega_1$, we obtain
\begin{equation}\label{react4_1}
\eta_{\alpha\beta}^{(1)} = \sum_n \theta (-\xi^+_n)
\langle\chi_n^+|\partial^2{\hat\xi}^+/\partial R_\alpha\partial R_\beta|\chi_n^+\rangle - \sum_n \theta (-\xi^-_n)
\langle\chi_n^-|\partial^2{\hat\xi}^-/\partial R_\alpha\partial R_\beta|\chi_n^-\rangle \,.
\end{equation}
The second contribution to $\eta_{\alpha\beta} (\tau - \tau')$ is (we keep only $\tau$ dependence in $G_0$ to shorten the notations):
\begin{equation}\label{react5}
\int d\tau d\tau^\prime \eta_{\alpha\beta}^{(2)}(\tau-\tau^\prime)=  Tr \big[G^s_0 (\tau, \tau') {\partial{\hat\xi^s}\over \partial R_\alpha (\tau)} G^s_0 (\tau', \tau) {\partial{\hat\xi}\over \partial R_\beta (\tau')}\big].
\end{equation}
where $s = \pm$.
In the Fourier representation
\beq\label{react51}
\eta_{\alpha\beta}^{(2)}(\omega) =  \sum_{n,m,s} \int {d\omega_1\over 2\pi}{\langle \chi_n^s |\partial{\hat\xi}^s/\partial R_\beta|\chi_m^s\rangle \langle \chi_m^s |\partial{\hat\xi}^s/\partial R_\alpha|\chi_n^s\rangle  \over (i(\omega_1+\omega) -s\xi_n^s)(i\omega_1 -s\xi_m^s)}\,.
\eeq
Performing the integration over  $\omega_1$, we obtain
\beq\label{react61}
\eta_{\alpha\beta}^{(2)}(\omega) =\sum_{n \neq m,s} {\langle \chi_n^s |\partial{\hat\xi}^s/\partial R_\beta|\chi_m^s\rangle \langle \chi_m^s |\partial{\hat\xi}^s/\partial R_\alpha|\chi_n^s\rangle  \over i\omega -s\xi_n^s +s\xi_m^s}\big[\theta(-s\xi_m^s)-\theta(-s\xi_n^s)\big].
\eeq
Let us now expand Eq. (\ref{react61}) in powers of $\omega$,
\begin{eqnarray}\label{react71}
\eta_{\alpha\beta}^{(2)}(\omega) = \eta_{\alpha\beta}^{(20)} + i\omega \eta_{\alpha\beta}^{(21)} + ...\,,
\end{eqnarray}
where
\begin{eqnarray}\label{react72}
\eta_{\alpha\beta}^{(20)} &=&  \sum_{n \neq m,s} {\langle \chi_n^s |\partial{\hat\xi}^s/\partial R_\beta|\chi_m^s\rangle \langle \chi_m^s |\partial{\hat\xi}^s/\partial R_\alpha|\chi_n^s\rangle  \over s\xi_m^s -s\xi_n^s}\big[\theta(-s\xi_m^s)-\theta(-s\xi_n^s)\big]  \,,\nonumber\\
\eta_{\alpha\beta}^{(21)} &=& - \sum_{n \neq m,s} {\langle \chi_n^s |\partial{\hat\xi}^s/\partial R_\beta|\chi_m^s\rangle \langle \chi_m^s |\partial{\hat\xi}^s/\partial R_\alpha|\chi_n^s\rangle  \over (\xi_m^s -\xi_n^s)^2}\big[\theta(-s\xi_m^s)-\theta(-s\xi_n^s)\big]  \,.
\end{eqnarray}
We will need the antisymmetric part of $\eta_{\alpha\beta}^{(21)}$ term.  Before evaluating it, we pause for a moment and  show that the frequency independent term
$\eta_{\alpha\beta}^{(20)}$ cancels $\eta_{\alpha\beta}^{(1)}$ from  Eq. (\ref{react4_1}).
This follows from the identity
\begin{equation}\label{react73}
\langle \chi_n^s |\partial{\hat\xi}^s/\partial R_\beta|\chi_m^s\rangle = (\xi_n^s - \xi_m^s)\langle \chi_n^s|\partial/\partial R_\beta|\chi_m^s\rangle,
\end{equation}
which is obtained from the condition $\partial/\partial R_\beta \langle \chi_n^s|{\hat\xi}^s |\chi_m^s\rangle  = \partial/\partial R_\beta (\xi_m^s \langle \chi_n^s|\chi_m^s\rangle) =0$, when $n \neq m$, by differentiating each term in $\partial/\partial R_\beta \langle \chi_n^s|{\hat\xi}^s |\chi_m^s\rangle$ over $R_{\beta}$.
Substituting this identity into (\ref{react72}) and using the completeness relation $\sum_n |\chi_n^s\rangle\langle\chi_n^s| =1$, we obtain for $\eta_{\alpha\beta}^{(20)}$ the same expression as in (\ref{react4_1}), but with the opposite sign.

We now return to $\eta_{\alpha\beta}^{(21)}$. Using the same identity and the completeness relation, we re-write  $\eta_{\alpha\beta}^{(21)}$ in Eq. (\ref{react72}) as
\begin{equation}\label{react74}
\eta_{\alpha\beta}^{(21)} =  \sum_n \theta(-\xi_n^+)\, \langle\chi_n^+|{\partial\over\partial R_\alpha}{\partial\over\partial R_\beta} - {\partial\over\partial R_\beta}{\partial\over\partial R_\alpha}|\chi_n^+\rangle - \sum_n \theta(-\xi_n^-)\, \langle\chi_n^-|{\partial\over\partial R_\alpha}{\partial\over\partial R_\beta} - {\partial\over\partial R_\beta}{\partial\over\partial R_\alpha}|\chi_n^-\rangle\,.
\end{equation}

To evaluate the r.h.s. of Eq. (\ref{react74}), we use the explicit form of $|\chi_n^s\rangle$. These are the eigenstates of $\hat\xi^s = -(\nabla + (is/2)\nabla\phi)^2/(2m)-\mu$. In polar coordinates $r$ and $\phi$, $\hat\xi^s$ can be written as
\begin{equation}\label{react8}
\hat\xi^s = -{1\over 2m}\Big[{\partial^2\over\partial r^2} + {1\over r}{\partial\over\partial r}  +{1\over r^2}\Big({\partial\over\partial\phi} + {is\over 2}\Big)^2 \Big] -\mu\,.
\end{equation}
The complete set of the eigenstatates of (\ref{react8}) is
\begin{equation}\label{react8_1}
\chi_n^\pm\equiv\chi_{\nu,k}^\pm(r,\phi)= \sqrt{k\over 2D}e^{i\nu\phi}J_{|\nu\pm 1/2|}(kr),
\end{equation}
where $\nu=\pm 1/2, \pm 3/2, ...$, and $k$ is  quasi-continuous radial wave number quantized as $\Delta k = \pi/D$, where, we remind, $2D$ is the system size (the quantization of $k$  originates from  the boundary condition $J_{|\nu\pm1/2|}(kD)=0$ at $kD\gg 1$).

Substituting these eigenstates and eigenvalues into Eq. (\ref{react74}) and using the fact  that for $\phi (r-R)$ given by (\ref{vortphase}),
\begin{equation}\label{www}
\Big({\partial\over\partial R_x}{\partial\over\partial R_y} - {\partial\over\partial R_y}{\partial\over\partial R_x}\Big)\phi = 2\pi\delta({\bf r}-{\bf R})\,,
\end{equation}
we obtain after straightforward algebra  $\eta_{\alpha \beta}^{(21)} = \epsilon_{\alpha\beta} {\bar \eta}$, where
\begin{equation}\label{react9}
{\bar \eta} = -2\pi \sum_{k,\nu=n+1/2}\theta\Big(\mu-{k^2\over 2m}\Big)\, \Big({k\over 2D}\Big) \nu \left(J^2_{|\nu+1/2|}(0) -J^2_{|\nu - 1/2|}(0)\right) \,.
\end{equation}
Because $J_n (0) =0$ for integer $n>0$ and  $J_0(0)=1$, the sum over $\nu$  gives $-1$. The summation over $k$
gives, at $D \to \infty$
\begin{equation}\label{react91}
\frac{\pi}{D} \sum_k k \theta\Big(\mu-{k^2\over 2m}\Big) ~= \int_0^\infty dk k \theta\Big(\mu-{k^2\over 2m}\Big) = m\mu = \pi n_0
\end{equation}
such that
\beq
\bar \eta = \pi n_0\,.
\eeq
Substituting into (\ref{exx_2}) we obtain the additional contribution to the Berry phase action of a moving vortex
\beq
S_{Berry,2} = S^0_{norm} ({\bf R}) - S^0_{norm} ({\bf R}=0) = i \pi A_{vort,2}\left(X(\tau) {\dot Y} (\tau) - Y(\tau) {\dot X} (\tau)\right)\,,
\label{exx_new}
\eeq
where
\beq
A_{vort,2}  =  -\frac{n_0}{2}.
\eeq
We emphasize that this term comes from the states right at the vortex core (see (\ref{www}), and in this respect is very different from
$A_{vort,1}$, which comes from the states far away from the vortex core.

\subsubsection{Effects of impurities}
We expect the result $A_{vort}=-(n_0)/2$ to hold  when the impurity potential is included. Indeed, Eq. (\ref{react74}) is valid when the impurity potential ($U_{imp}$) is present and  the eigenfunctions $\chi^\pm_n\rangle$ can still be expressed as
\beq
\chi^\pm_n ({\bf r}) = e^{\pm i \phi({\bf r}-{\bf R})/2} a_n({\bf r})\,,
\label{diss1}
\eeq
where $a({\bf r})$ satisfies the Schrodinger equation
\beq
\big[-\nabla^2/(2m) + U_{imp}(({\bf r}))\big] a_n({\bf r}) = \xi_n a_n({\bf r})  \,.
\label{diss2}
\eeq
Substituting this $ \chi^\pm_n ({\bf r})$   into Eq. (\ref{react74}), we obtain
\beq
\eta_{\alpha\beta}^{(21)} =  \pi \sum_n \theta(-\xi_n)\, |a_n({\bf R})|^2  + \pi \sum_n \theta(-\xi_n)\, |a_n({\bf R})|^2 = \pi n_0({\bf R})  \,,
\label{diss3}
\eeq
where $n_0({\bf R})$ is the fermion density at the vortex core, i.e., the same result as in the absence of impurity potential.

\subsection{The total Berry phase term for a moving vortex}.

Combining the two contributions to $S_{Berry}$, we obtain the total Berry phase term for a moving vortex
\beq
S^{vort}_{Berry} =
S_{Berry,1} + S_{Berry,2 }  =   i\pi A_{vort}
\left(X(\tau) {\dot Y} (\tau) - Y(\tau) {\dot X} (\tau)\right)
\label{tot_berry}
\eeq
where
\beq
A_{vort} = \frac{n- n_0}{2}.
\label{last}
\eeq
We remind that $n-n_0 = 2 N_0 E_0$ when $E_0 < E_F$, and $n-n_0 = 2 N_0 E_F$ when $E_0 > E_F$, where $2E_0$ is the bound state energy of two fermions in a vacuum.  The first limit corresponds to BCS, the second one to BEC. More specifically, $n_0 =0$ when $E_0 > E_F$, hence in this situation the prefactor in the Berry phase term in the action becomes just $i \pi n/2$.    The vanishing of $n_0$ once $\mu$ becomes negative is consistent with the generic reasoning in Ref.~\cite{volovik} that free-fermion contribution to $A_{vort}$ vanishes once the system undergoes a (fictitious) Lifshitz transition, in which the (fictitious) Fermi surface of free fermions with renormalized $\mu$ disappears. In our case, this happens once $E_0$ becomes larger than $E_F$.

We also note that the two contributions to $S_{Berry}$ from $\Delta \to 0$ -- one given by Eq. (\ref{exx_new}) and the other by Eq. (\ref{Bvortfin_1}), are equal in magnitude, but differ in sign.  As a result, the combined total contribution from the action at $\Delta \to 0$  vanishes. As the consequence, and the total Berry phase term in the action of a moving vortex is the same as in Eq. (\ref{chna_10}), obtained by expanding in $\Delta$.

We argued above that $A_{vort,2} =-n_0/2$ is not affected by impurities, i.e., the reaction force remains the same in the presence of imputity potential. By the same reason,  Eq. (\ref{S1}) and  the subsequent consideration in Sec. IV for the Magnus force also remains valid when the impurity potential is present. This is consistent with argument made by  Ao and Thouless\cite{thouless} that  impurity scattering should not modify the value of the Magnus force.  As the consequence, we expect  $A_{vort}=(n-n_0)/2$ to hold when impurity scattering is present.

\subsubsection{The interpretation of Eq. (\ref{tot_berry})}

As we said, there are two contributions to the prefactor $A_{vort}$ for the Berry phase term in the effective action for a moving vortex -- one, $A_{vorx,1}$,  comes from states far away from a vortex core, and the other, $A_{vort,2}$, comes from the states at the vortex core. Looking back at our derivation of  $A_{vort,1}$, we see that this term has two contributions: one comes from the first term in the action in Eq. (\ref{Z}), another comes from $\dot \phi$ piece in last term, $S_{norm}$,  which is the normal state contribution to the action (more accurately, the contribution from $\Delta \to 0$).  On more careful look, we note that there are in fact two contributions from $\Delta \to 0$ in $A_{vort,1}$: the one from $S_{norm}$ and the one from the lower limit in the integral  $\int_0^1 d\lambda$ in the first term in (\ref{Z}) (the lower limit $\lambda =0$ corresponds to $\Delta \to 0$).  We separated the two contributions for convenience of the derivation and to show explicitly how parasitic ultra-violet divergent term  $\sum_k (1)$ cancels out between the contributions from $\lambda =1$ and $\lambda=0$ (see Eqs. (\ref{aa6}) and (\ref{aa6_1})). If we were to combine from the start the $\dot \phi$ piece in $S_{norm}$ and the contribution from $\lambda =0$ in the first term in (\ref{Z}), we would obtain that the $n_0$ terms cancel each other (one is $-n_0/2$, another is $n_0/2$). The cancellation implies that there is no contribution from $\Delta \to 0$ to the Berry phase term besides the counter-term to cancel the ultra-violet divergence. The  full $A_{vort,1} = n/2$ comes exclusively from the limit $\lambda =1$ in the first term in (\ref{Z}), which  describes the action at  $\Delta ({\bf r}, \tau, 1) = \Delta ({\bf r}, \tau)$, where $\Delta ({\bf r}, \tau)$ is the actual gap function at distance ${\bf r}$ from a vortex.  We recall that this contribution comes from fermionic states far away from a vortex core, where the gap amplitude approaches equilibrium value $\Delta_0$. Obviously then, $A_{vort,1} =n/2$ is the same as the prefactor for $\dot \phi$ term in the effective action for the case when $\phi ({\bf r}, \tau)$ is a regular function of its arguments.

The second contribution $A_{vort,2}$ is additional contribution from the vortex core, i.e., from ${\bf r} =0$.  In general,  low-energy fermionic states near the core, the ones  with energies below $\Delta_0$, are discrete levels with separation $\omega_0 \sim \Delta^2_0/E_F$ (Refs. \cite{bardeen,caroli,kopninvolovik}).  We found no contribution from discrete levels in the vortex core from the part of the action with the actual $\Delta ({\bf r}, \tau)$ (this would be a contribution from the upper limit $\lambda =1$ in the first term in (\ref{Z})). Our $A_{vort,2}$  comes from the $\nabla \phi$ piece in $S_{norm} = Tr \log(G_0^{-1})$, which is the contribution to the action from $\Delta \to 0$.  In this limit, the  spacing between discrete levels $\omega_0 \sim \Delta^2_0/E_F$ vanishes, and electronic states in the vortex core are 
not quantized and are
described by  a continuous variable $k$.

The Berry phase term in the effective action has been analyzed earlier. Several authors ~\cite{volovik,volovik_1,otterlo, otterlo_2,kopninvolovik} argued on general grounds  that quantization of fermionic states inside the vortex core can be neglected in the hydrodynamic limit $\omega_0 \tau \ll 1$, where  $\tau$ is fermionic lifetime.  In this limit, earlier works~~\cite{volovik,volovik_1,otterlo, otterlo_2,kopninvolovik} found the same 
  $A_{vort} =  (n-n_0)/2$ as in  Eq. (\ref{last}).  
 In our consideration, $\omega_0 =0$ and $\tau = \infty$, so $\omega_0 \tau$ is not well determined.  Still, 
 our $A_{vort,2}$ comes from continuous (i.e., non-quantized) states, and 
   it remains the same  in the presence of impurities, i.e., at a finite $\tau$. In this respect, we believe that the agreement between our $A_{vort}$ and the one obtained in earlier works at $\omega_0 \tau \ll 1$ is meaningful.
   

There is, however, one aspect in which our result seems to differ from earlier works. Namely, these works speculated~\cite{volovik,volovik_1,otterlo, otterlo_2,kopninvolovik}
 that at finite $\omega_0 \tau$  there should be a contribution to $A_{vort}$ from discrete levels in the vortex core. These and several other authors have argued ~\cite{volovik,volovik_1,kopninvolovik,volovik_book,otterlo_2,stone} that in the limit $\omega \tau \gg 1$, the total contribution from the vortex core $A_{vort,2}$ should vanish, i.e., the total $A_{vort}$ should reduce to  $A_{vort,1} = n/2$. 
   We  didn't find in our microscopic approach the contribution to $A_{vort}$ from  discrete levels in the vortex core in the 
    term in the action with a finite $\Delta ({\bf r}, \tau)$.
   It remains to be seen whether such contribution can be obtained by going beyond the approximations we made in 
    our  derivation of the effective action.
  

\subsection{External superflow and the equation for the balance of forces}

In the presence of an external supercurrent,  the phase of the order parameter in Eq. (\ref{vort}) acquires an additional term, $2m{\bf v}_s{\bf r}$.  The effect of this term on the action can be analyzed perturbatively if  ${\bf v}_s$ is small. Performing the same  gauge transformation as we used to move from  Eq. (\ref{bdgmain}) to Eq. (\ref{ham}), but for non-zero ${\bf v}_s$, we obtain the additional -$i{\bf v}_s\cdot\nabla$ term in the lhs of Eq. (\ref{ham}) (Ref. \cite{com3}).  Evaluating now the correction to the ground state energy within the first order perturbation theory, we obtain
\begin{equation}\label{V'}
\delta E(v_s) = -i\sum_n {\bf v}_s \int d^2{\bf r} \langle\chi_{n}(\tau,{\bf r})|\nabla\chi_{n}(\tau, {\bf r})\rangle  \theta(-E_n)   \,,
\end{equation}
where $|\chi_{n}(\tau, {\bf r})\rangle$ are the solutions to the BdG equations in the presence of a vortex, but without ${\bf v}_s$, see Eq. (\ref{chibound}).  From (\ref{V'}) we then obtain the extra term in the action of the vortex:
\begin{equation}\label{S'}
{\cal S}_{v_s} = -i\int_{-\infty}^\infty d\tau \sum_n\,\int d^2{\bf r}\,{\bf v}_s \Big[\langle\chi_{n}(\tau,{\bf r})|\nabla\chi_{n}(\tau, {\bf r})\rangle -  \langle\chi^{|\Delta|\rightarrow 0}_{n}(\tau,{\bf r})|\nabla\chi^{|\Delta|\rightarrow 0}_{n}(\tau, {\bf r})\rangle  \Big]  \theta(-E_n)\,.
\end{equation}
The sum and the matrix elements in the rhs of Eq. (\ref{S'}) are identical to those in
Eq. (\ref{S1}), hence
\begin{equation}\label{S'1}
{\cal S}_{v_s} = -i \pi B_{vort}
\int d \tau  \left( X(\tau) {v_{sy}}  -  Y(\tau) {v_{sx}} \right)\,,
\end{equation}
where $B_{vort} = A$, with $A$ given in Eq. (\ref{chna_10}), e.g. $A = (n-n_0)/2$ in the BCS regime, and  $A=n/2$ in the BEC regime.

As we have seen in the previous subsection, there exist another contribution to the force acting on a vortex associated with the vortex core. Similarly, one might expect that there is another contribution to the the action in Eq. (\ref{S'1}), i.e., to the constant $B_{vort}$ that comes from the core. As we show below, this is not the case, - we find that the normal part of the action, $S_{norm}$ does not contribute to $B_{vort}$.

To see this, let's look at $S_{norm}$ in Eqs. (\ref{ss_211}), (\ref{ss_21}) in the presence of extra phase superfluid velocity. We need to replace in this equation $\nabla\phi$ by $\nabla\phi+2m {\bf v}_s$. This  gives an extra term under the $Tr\log{...}$ (Ref. \cite{new_comm}),
\beq
{\hat V}_s = -i{\bf v}_s(\nabla + {i\over 2}\sigma_z \nabla\phi).
\label{vsop}
\eeq
Treating this extra term perturbatively, we expand the $Tr \log{(...)}$ to the first order in $v_s$. The term $Tr(G_0{\hat V}_s)$ does not contribute because the expectation values of $\nabla$ and of $i\sigma_z \nabla\phi/2$ in Eq. (\ref{vsop}) cancel each other. The next-order term $Tr[G_0{\hat V}_sG_0(\partial{\hat\xi}/\partial R_\beta)R_\beta]$ is apparently relevant as it contains the same combination $v_{s\alpha}R_\beta$ as in (\ref{S'1}). Using the same manipulations as in the previous subsection, we can write
\beq
Tr[G_0{\hat V}_sG_0(\partial{\hat\xi}/\partial R_\alpha)R_\alpha] = -i\int d\tau \, B_{core}^{\alpha, \beta}\, v_{s\alpha}R_\beta(\tau) \,,
\label{vsex}
\eeq
where
\bea
B_{core}^{\alpha, \beta} &=& \sum_{n \neq m,s} {\langle \chi_n^s |[\nabla_\alpha + (is/2)(\nabla_\alpha\phi)]|\chi_m^s\rangle \langle \chi_m^s |\partial{\hat\xi}^s/\partial R_\beta|\chi_n^s\rangle  \over \xi_n^s -\xi_n^s}\big[\theta(-\xi_m^s)-\theta(-\xi_n^s)\big] \nonumber\\
&=& \sum_{n, s} \langle \chi_n^s |\Big[\nabla_\alpha + {is\over 2}(\nabla_\alpha\phi),\ {\partial\over \partial R_\beta}\Big]|\chi_n^s \rangle\,\theta(-\xi_n^s) \,.
\label{vsex1}
\eea

The term $(is/2)(\nabla_\alpha\phi)$ in Eq. (\ref{vsex1}) doe not contribute: Since $[(\nabla_\alpha\phi),\,\partial/\partial R_\beta]= -\partial^2\phi/\partial_\alpha\partial_\beta$, it cancels a correction to $Tr [G_0 i\sigma_z {\bf v}_s\nabla\phi/2]$ that arises when we expand $\nabla\phi$ as $\nabla_\alpha\phi(0)+ \nabla_\alpha\nabla_\beta\phi(0)R_\beta +...$, i.e., the $Tr [G_0 i\sigma_z v_{s\alpha} \nabla_\alpha\nabla_\beta\phi(0)R_\beta/2]$ term. This is similar to cancellation between Eqs. (\ref{react4_1}, \ref{react72}) in the previous subsection. For the  remaining $\nabla_\alpha$ piece in Eq. (\ref{vsex1}) we obtain
\bea
B_{core}^{\alpha, \beta} = \sum_n \theta(-\xi_n^+)\, \langle\chi_n^+|{\partial\over\partial R_\alpha}{\partial\over\partial R_\beta} - {\partial\over\partial R_\beta}{\partial\over\partial R_\alpha}|\chi_n^+\rangle + \sum_n \theta(-\xi_n^-)\, \langle\chi_n^-|{\partial\over\partial R_\alpha}{\partial\over\partial R_\beta} - {\partial\over\partial R_\beta}{\partial\over\partial R_\alpha}|\chi_n^-\rangle\,.
\label{vsex2}
\eea
The expression for $B_{core}$ in Eq. (\ref{vsex2}) is similar to the formula for $\eta_{\alpha\beta}^{(21)}$ in Eq. (\ref{react74}), but with one important distinction: the relative sign between the two terms in
(\ref{vsex2})  is plus, while in (\ref{react74}) it is  minus. As a result, performing the same calculations as in Eqs. (\ref{react8} - \ref{react9}), we obtain
\bea
B_{core}^{\alpha, \beta} &=& 2\pi \sum_{k,\nu=n+1/2}\theta\Big(\mu-{k^2\over 2m}\Big)\, \Big({k\over 2D}\Big) \nu \left(J^2_{|\nu+1/2|}(0) +J^2_{|\nu - 1/2|}(0)\right)\nonumber\\
&=& 4\pi \sum_{k,n}\theta\Big(\mu-{k^2\over 2m}\Big)\, \Big({k\over 2D}\Big) n\,J^2_{|n|}(0)\,.
\label{vsex3}
\eea
Because the product $nJ^2_{|n|}(0)$ is zero for any integer $n$, $B_{core}^{\alpha, \beta}=0$. Hence,  $S_{norm}$ does not contribute to $B_{vort}$ in Eq. (\ref{S'1}).

Eq. (\ref{S'1}), together with Eq. (\ref{Bvortfin}), determines the balance of forces acting on a vortex.
Converting from Matsubata to real time,  we obtain
\begin{equation}\label{Fbal}
A_{vort} \dot{\bf R}\times {\bf z} - B_{vort} {\bf v}_s\times {\bf z}=0  \,,
\end{equation}
where ${\bf z}$ is a unit vector perpendicular to the 2d plane.
We see from Eq. (\ref{Fbal}) that
\beq
v_{vort} = {\dot {\bf R}} = \frac{B_{vort}}{A_{vort}}~ {\bf v}_s.
\label{yyy1}
\eeq
Because $A_{vort} = B_{vort} = (n-n_0)/2$, we have
\beq
{\bf v}_{vort} = {\bf v}_{s}.
\label{yyy2}
\eeq
This agrees with the reasoning based on translational invariance~\cite{volovik,volovik_book,stone}.

\section{Summary}
In this paper we  analyzed the evolution of the $T=0$ low frequency dynamics of collective excitations of an s-wave neutral superconductor between BCS and BEC regimes.  The two regimes correspond to small and large ratio of $E_0/E_F$, respectively, where $E_F$ is the Fermi energy, and $E_0$ is the bound state energy for two particles. In D=2, bound state develops already at weak coupling, what allows one to analyze the crossover within a controllable weak coupling expansion. We obtained the terms in the long-wavelength action, proportional to $(\nabla \phi)^2$ and ${\dot \phi}^2$, where $\phi ({\bf r}, t)$ is the  phase of the superconducting  order parameter $\Delta ({\bf r}, t) = \Delta e^{i\phi ({\bf r}, t)}$. We  found that the phase velocity of the collective excitations remains $v_F/\sqrt{2}$ through the BCS-BEC crossover. We also obtained the topological Berry phase term in the long-wavelength  action $ i\int d \tau A {\dot \phi}$. We found that the prefactor  $A = n/2$, where $n$ is the  actual fermion density, and does not change through BCS-BEC crossover.

The  Berry phase term in the action is meaningful when the phase of the superconducting order parameter is not defined globally, which is the case when the pairing gap vanishes at some point is space, like in the vortex core.  We computed the effective action for a moving vortex in a neutral s-wave superconductor in 2d. The Berry phase term for a moving vortex  has the form $S_{Berry} = i \pi A_{vort} \int d \tau  \left( X(\tau) {\dot Y (\tau)}  -  Y(\tau) {\dot X (\tau)}\right)$, where $X (\tau)$ and $Y (\tau)$ are coordinates of the center of a moving vortex. We found that two contributions to the prefactor $A_{vort}$. One comes from fermionic states far away from the vortex core and is the same as in the long-wavelength action -- $A_{vort,1} = n/2$. Another comes from  delocalized (continuous) fermionic states inside the vortex core. For this second contribution we obtained $A_{vort,1} = -n_0/2$, where $n_0$ is the fermionic density at the vortex core (it is equal to the fermionic density in the normal state, but for the same chemical potential $\mu$ as in the superconducting state). In physical terms, the long-wavelength contribution $n/2$ represents a Magnus force acting on a moving vortex, while the offset term $-n_0/2$ represents a reaction force from normal fermions at the vortex core. The total  $A_{vort}=(n-n_0)/2$. In the BCS limit, $n-n_0 \ll n$, i.e., the total transverse Lorentz-like force acting on the vortex is much smaller than the Magnus force.  In the BEC limit, $n_0 =0$ because the effective $\mu <0$, and there are no normal state fermions at the vortex core. Then one recovers the result that the total transverse force equals  to the Magnus force. We argued that the result for $A_{vort}$ remains valid in the presence of impurity scattering. Finally, we found that in the presence of an external superflow the vortex dynamics obeys Galilean (translational) invariance principle: The vortices move together with the superflow.

The result $A_{vort} = (n-n_0)/2$ agrees with  earlier works ~\cite{volovik,volovik_1,otterlo_2,kopninvolovik}, which obtained this $A_{vort}$ neglecting the quantization of fermionic states in the vortex core. In agreement with these results, we found that in our approach the contribution from the states near the vortex core comes only from the part of the action at $\Delta \to 0$, when the spacing between the states in the vortex core vanishes, and the low-energy states become continuous. Earlier works ~\cite{stone,volovik,otterlo,volovik_book,kopnonvolovik,otterlo_2}speculated that there should be another contribution to  $A_{vort}$ from discrete states in the vortex core. We didn't find such contribution in our analysis of the effective action.  This term may emerge once one moves beyond our approach, based on the evaluation of the effective action for the vortex motion.

Our results for the expansion of the effective action in terms of time derivatives of slowly varying order parameter (Eqs. (\ref{S0}), (\ref{S1}), and (\ref{S2})) can be straightforwardly extended to other symmetries of the order parameter and to non-Galilean-invariant  dispersion, as long as adiabatic approximation is applicable.  We note in this regard that the  topological term in the action  plays a special role in superconductors with the nodes in the order parameter, e.g., it determines the magnitude of the orbital momentum in the A-phase of a p-wave superconductor, like $^3He-A$ (Ref. \cite{A-phase}). The terms of higher orders in powers of $\partial_\tau\Delta$  or in higher derivatives of $\Delta$ can be obtained from Eqs. (\ref{Delta}) and ( \ref{G}), though in practice such calculation is likely to be rather cumbersome.

\section{Acknowledgements}
The authors thank I. Aleiner, P. Ao, J. Dziarmaga, I. Martin, J. Sauls, D. Solenov, and G.E.  Volovik for valuable discussions. We are  particularly thankful to M. Stone for in-depth discussions (and respectful disagreement), highly relevant comments and suggestions,  and for pointing out an error in the earlier version of the MS.

The work by DM was supported by the U.S. Department of Energy through the Los Alamos National Laboratory. Los Alamos National Laboratory is operated by Triad National Security, LLC, for the National Nuclear Security Administration of the U.S. Department of Energy (Contract No. 89233218NCA000001). The work by AVC  was supported by the Office of Basic Energy Sciences U. S. Department of Energy under award DE-SC0014402.  AVC is thankful to KITP at UCSB  where part of the work has been done. KITP is supported by NSF grant PHY-1125915.

\appendix
\section{Derivation of the action to second order in time derivative}
Substituting Eq. (\ref{G2}) into Eq. (\ref{F0}) we get two contributions,
\begin{equation}\label{F2}
L^{(2)}= L_1^{(2)} + L_2^{(2)}\equiv \int_0^1 d\lambda \big[l_1^{(2)}(\lambda,\tau)+ l_2^{(2)}(\lambda,\tau) \big],
\end{equation}
each generated by first and second terms in the rhs of Eq. (\ref{G2}), respectively:
\begin{eqnarray}\label{F20}
l^{(2)}_1(\lambda,\tau)=\int {d\omega\over 2\pi} e^{i\omega\epsilon^+}\sum_{n,l}{\langle\chi_{l,\lambda}|{\partial_\lambda\hat \Delta}|\chi_{n,\lambda}\rangle\langle \chi_{n,\lambda}|\partial_\tau^2{\hat \Delta}(\tau)|\chi_{l,\lambda}\rangle\over (i\omega - E_{n,\lambda})^3(i\omega - E_{l,\lambda})}\,,  \\\label{F21}
l^{(2)}_2(\lambda,\tau)=\int {d\omega\over 2\pi} e^{i\omega\epsilon^+}\sum_{n,m,l} {\langle\chi_{l,\lambda}|\partial_\lambda{\hat \Delta}|\chi_{m,\lambda}\rangle\langle \chi_{m,\lambda}|\partial_\tau{\hat \Delta}(\tau)|\chi_{n,\lambda}\rangle\langle \chi_{n,\lambda}|\partial_\tau{\hat \Delta}(\tau)|\chi_{l,\lambda}\rangle\over (i\omega - E_{n,\lambda})(i\omega - E_{l,\lambda})^2(i\omega - E_{m,\lambda})^2}    \,.
\end{eqnarray}

The $\omega$-integration Eq. (\ref{F20}) gives
\begin{eqnarray}\label{F2fin1}
l_1^{(2)}(\lambda,\tau)  = \sum_{n,l}{\langle\chi_{l,\lambda}|\partial_\lambda{\hat \Delta}|\chi_{n,\lambda}\rangle\langle \chi_{n,\lambda}|\partial_\tau^2{\hat \Delta}(\tau)|\chi_{l,\lambda}\rangle (\theta_n -\theta_l) \over (E_{n,\lambda} - E_{l,\lambda})^3} = \sum_{n,l}{\langle\partial_\lambda\chi_{l,\lambda}|\chi_{n,\lambda}\rangle\langle \chi_{n,\lambda}|\partial_\tau^2{\hat H}|\chi_{l,\lambda}\rangle (\theta_n -\theta_l) \over (E_{n,\lambda} - E_{l,\lambda})^2} \,,
\end{eqnarray}
where in the second equality we have used Eq. (\ref{hfoff}).

The second integral, e.g. Eq. (\ref{F20}) is less straightforward. Let us exclude terms with repeated indices from the the triple sum in Eq. (\ref{F21}). Then the integrand has three poles and, after some algebra, we find (the situation when two indices coincide will be considered separately below),
\begin{eqnarray}\label{F2fin2}
l_2^{(2)}(\lambda,\tau)  = \sum_{n,m,l}{\langle\partial_\lambda\chi_{l,\lambda}|\chi_{m,\lambda}\rangle\langle\partial_\tau\chi_{m,\lambda}|\chi_{n,\lambda}\rangle \langle\partial_\tau\chi_{n,\lambda}|\chi_{l,\lambda}\rangle (E_{l,\lambda} - E_{m,\lambda}) \theta_n  \over (E_{n,\lambda} - E_{l,\lambda})(E_{m,\lambda} - E_{n,\lambda})} \\\nonumber
+\sum_{n,m,l}{\langle\partial_\lambda\chi_{l,\lambda}|\chi_{m,\lambda}\rangle\langle\partial_\tau\chi_{m,\lambda}|\chi_{n,\lambda}\rangle \langle\partial_\tau\chi_{n,\lambda}|\chi_{l,\lambda}\rangle (E_{n,\lambda} - E_{l,\lambda})(2E_{n,\lambda}-3E_{m,\lambda} + E_{l,\lambda}) \theta_m  \over (E_{n,\lambda} - E_{m,\lambda})(E_{m,\lambda} - E_{l,\lambda})^2}\\\nonumber
+\sum_{n,m,l}{\langle\partial_\lambda\chi_{l,\lambda}|\chi_{m,\lambda}\rangle\langle\partial_\tau\chi_{m,\lambda}|\chi_{n,\lambda}\rangle \langle\partial_\tau\chi_{n,\lambda}|\chi_{l,\lambda}\rangle (E_{m,\lambda} - E_{n,\lambda})(2E_{n,\lambda}-3E_{l,\lambda} + E_{m,\lambda}) \theta_l  \over (E_{n,\lambda} - E_{l,\lambda})(E_{m,\lambda} - E_{l,\lambda})^2}\ ,
\end{eqnarray}
where we again used Eqs. (\ref{hfoff}, \ref{hfofftau}).

Next we transform Eq. (\ref{F2fin1}) by using
\begin{equation}\nonumber
\langle \chi_{l,\lambda}|\partial_\tau^2{\hat \Delta}(\tau)|\chi_{n,\lambda}\rangle = \partial_\tau\langle \chi_{l,\lambda}|\partial_\tau{\hat \Delta}(\tau)|\chi_{n,\lambda}\rangle - \langle \partial_\tau\chi_{l,\lambda}|\partial_\tau{\hat \Delta}(\tau)|\chi_{n,\lambda}\rangle - \langle \chi_{l,\lambda}|\partial_\tau{\hat \Delta}(\tau)|\partial_\tau\chi_{n,\lambda}\rangle \,,
\end{equation}
and integrating by parts the term containing $\partial_\tau\langle \chi_{l,\lambda}|\partial_\tau{\hat \Delta}(\tau)|\chi_{n,\lambda}\rangle$ (we recall that $l_1^{(2)}$ is under $\tau$ integration when substituted in Eqs. (\ref{Z}, \ref{F})):
\begin{eqnarray}\label{F2fin12}
l_1^{(2)}(\lambda,\tau)  = \sum_{n,l}\Big\{2{\langle\partial_\lambda\chi_{l,\lambda}|\chi_{n,\lambda}\rangle\langle \partial_\tau\chi_{n,\lambda}|\chi_{l,\lambda}\rangle\partial_\tau (E_{n,\lambda} - E_{l,\lambda}) \over (E_{n,\lambda} - E_{l,\lambda})^2} -{\langle\partial_\lambda\partial_\tau\chi_{l,\lambda}|\chi_{n,\lambda}\rangle\langle \partial_\tau\chi_{n,\lambda}|\chi_{l,\lambda}\rangle \over E_{n,\lambda} - E_{l,\lambda}}\ \ \ \ \ \ \ \ \ \\\nonumber
-{\langle\partial_\lambda\chi_{l,\lambda}|\partial_\tau\chi_{n,\lambda}\rangle\langle \partial_\tau\chi_{n,\lambda}|\chi_{l,\lambda}\rangle \over E_{n,\lambda} - E_{l,\lambda}}-{\langle\partial_\lambda\chi_{l,\lambda}|\chi_{n,\lambda}\rangle\langle\partial_\tau \chi_{n,\lambda}|\partial_\tau{\hat H}|\chi_{l,\lambda}\rangle \over (E_{n,\lambda} - E_{l,\lambda})^2}-{\langle\partial_\lambda\chi_{l,\lambda}|\chi_{n,\lambda}\rangle\langle \chi_{n,\lambda}|\partial_\tau{\hat H}|\partial_\tau\chi_{l,\lambda}\rangle \over (E_{n,\lambda} - E_{l,\lambda})^2}\Big\}(\theta_n-\theta_l)\,.
\end{eqnarray}

The last three terms in the rhs of Eq. (\ref{F2fin12}) can be combined with Eq. (\ref{F2fin2}) when we insert $\sum_m|\chi_{m,\lambda}\rangle\langle \chi_{m,\lambda}|$ in the matrix elements in the terms that contain double derivatives (with respect to $\tau$ in the fourth and fifth terms and with respect $\lambda$ and $\tau$ in the third term). In doing so we treat separately the terms with $m\neq n$ and $m\neq l$ and the terms with $m=n$ and $m=l$.   Using  the relation
\begin{equation}\label{hftau}
\langle \chi_{m,\lambda}|\partial_\tau {\hat H}|\chi_{m,\lambda}\rangle = \partial_\tau \langle \chi_{m,\lambda}|{\hat H}|\chi_{m,\lambda}\rangle = \partial_\tau E_{m,\lambda}  \,,
\end{equation}
for the diagonal matrix elements, as well as $\langle \partial_\tau\chi_{m,\lambda}|\chi_{m,\lambda}\rangle=-\langle \chi_{m,\lambda}|\partial_\tau\chi_{m,\lambda}\rangle$, we obtain after some algebra
\begin{eqnarray}\label{F2fin13}
({\rm Last\ 3\ terms\ of\ Eq.\ A6})=\sum_{n,m,l}\langle\partial_\lambda\chi_{l,\lambda}|\chi_{m,\lambda}\rangle\langle\partial_\tau\chi_{m,\lambda}|\chi_{n,\lambda}\rangle \langle\partial_\tau\chi_{n,\lambda}|\chi_{l,\lambda}\rangle \ \ \ \ \ \ \ \ \ \ \ \ \ \ \ \ \ \ \ \ \\\nonumber
\times\Big[{\theta_n-\theta_l\over(E_{n,\lambda} - E_{l,\lambda})} - {(2E_{n,\lambda} - E_{l,\lambda}- E_{m,\lambda})(\theta_m-\theta_l)\over(E_{m,\lambda} - E_{l,\lambda})^2} \Big] + \sum_{n,l}{\langle\partial_\lambda\chi_{l,\lambda}|\chi_{n,\lambda}\rangle\langle \partial_\tau\chi_{n,\lambda}|\chi_{l,\lambda}\rangle\partial_\tau (E_{n,\lambda} - E_{l,\lambda})(\theta_n-\theta_l) \over (E_{n,\lambda} - E_{l,\lambda})^2}\\\nonumber
-\sum_{n,l}{\langle \partial_\tau\chi_{n,\lambda}|\chi_{l,\lambda}\rangle\langle\partial_\tau\chi_{l,\lambda}|\chi_{n,\lambda}\rangle\langle\partial_\lambda\chi_{l,\lambda}|
\chi_{l,\lambda}\rangle + \langle \partial_\tau\chi_{n,\lambda}|\chi_{l,\lambda}\rangle\langle\partial_\lambda\chi_{l,\lambda}|\chi_{n,\lambda}\rangle\langle\partial_\tau\chi_{l,\lambda}|
\chi_{l,\lambda}\rangle \over E_{n,\lambda} - E_{l,\lambda}}(\theta_n-\theta_l) \, .
\end{eqnarray}
In the triple sum, i.e. in the first term in the rhs of Eq. (\ref{F2fin13}), $m\neq n\neq l$, while the last term in Eq. (\ref{F2fin13}) arises due to $m=n$ and $m=l$ terms. Adding this triple sum with $l_2^{(2)}(\lambda,\tau)$ from Eq. (\ref{F2fin2}),   we obtain
\begin{eqnarray}\label{triplesum}
\sum_{n,m,l}{\langle\partial_\lambda\chi_{l,\lambda}|\chi_{m,\lambda}\rangle\langle\partial_\tau\chi_{m,\lambda}|\chi_{n,\lambda}\rangle \langle\partial_\tau\chi_{n,\lambda}|\chi_{l,\lambda}\rangle \over E_{m,\lambda} - E_{n,\lambda}}(\theta_m-\theta_n)\,.
\end{eqnarray}
The triple sum in Eq. (\ref{triplesum}) can be transformed into the double sum by using a completeness relation, $\sum_l|\chi_{l,\lambda}\rangle\langle \chi_{l,\lambda}|=1$,
\begin{eqnarray}\label{F2fin24}
-\sum_{n,m}{\langle\partial_\tau\chi_{n,\lambda}|\partial_\lambda\chi_{m,\lambda}\rangle\langle\partial_\tau\chi_{m,\lambda}|\chi_{n,\lambda}\rangle  \over E_{m,\lambda} - E_{n,\lambda}}(\theta_m-\theta_n)  +\sum_{n,m}{\langle\chi_{m,\lambda}|\partial_\lambda\chi_{m,\lambda}\rangle\langle\partial_\tau\chi_{m,\lambda}|\chi_{n,\lambda}\rangle\langle\partial_\tau\chi_{n,\lambda}|
\chi_{m,\lambda}\rangle   \over E_{m,\lambda} - E_{n,\lambda}}(\theta_m-\theta_n)   \\\nonumber
    +\sum_{n,m}{\langle\chi_{n,\lambda}|\partial_\lambda\chi_{m,\lambda}\rangle\langle\partial_\tau\chi_{m,\lambda}|\chi_{n,\lambda}\rangle\langle\partial_\tau\chi_{n,\lambda}|
\chi_{n,\lambda}\rangle   \over E_{m,\lambda} - E_{n,\lambda}}(\theta_m-\theta_n)
\ ,
\end{eqnarray}
where the last two terms in Eq. (\ref{F2fin24}) correspond to $l=n$ and $l=m$ terms, which are omitted in the triple sum in Eq. (\ref{triplesum}).  The last two terms in Eq. (\ref{F2fin24}) cancel the last term in Eq. (\ref{F2fin13}). Using this, we finally obtain that
\begin{eqnarray}\label{F2fin241}
({\rm Last\ 3\ terms\ of\ Eq.\ A6}) + l_2^{(2)}= -\sum_{n,m}{\langle\partial_\tau\chi_{n,\lambda}|\partial_\lambda\chi_{m,\lambda}\rangle\langle\partial_\tau\chi_{m,\lambda}|\chi_{n,\lambda}\rangle  \over E_{m,\lambda} - E_{n,\lambda}}(\theta_m-\theta_n)\\\nonumber
+\sum_{n,l}{\langle\partial_\lambda\chi_{l,\lambda}|\chi_{n,\lambda}\rangle\langle \partial_\tau\chi_{n,\lambda}|\chi_{l,\lambda}\rangle\partial_\tau (E_{n,\lambda} - E_{l,\lambda})(\theta_n-\theta_l) \over (E_{n,\lambda} - E_{l,\lambda})^2}\ .
\end{eqnarray}
The first term in the rhs of Eq. (\ref{F2fin241}) can be combined with the second term in Eq. (\ref{F2fin12}) as
\begin{eqnarray}\label{F2fin25}
-\sum_{n,m}{\langle\partial_\lambda\partial_\tau\chi_{n,\lambda}|\chi_{m,\lambda}\rangle\langle \partial_\tau\chi_{m,\lambda}|\chi_{n,\lambda}\rangle +  \langle\partial_\tau\chi_{n,\lambda}|\partial_\lambda\chi_{m,\lambda}\rangle\langle\partial_\tau\chi_{m,\lambda}|\chi_{n,\lambda}\rangle  \over E_{m,\lambda} - E_{n,\lambda}}(\theta_m-\theta_n)\\\nonumber
= -\sum_{n,m} {\big(\partial_\lambda\langle\partial_\tau\chi_{n,\lambda}|\chi_{m,\lambda}\rangle\big) \langle\partial_\tau\chi_{m,\lambda}|\chi_{n,\lambda}\rangle  \over E_{m,\lambda} - E_{n,\lambda}}(\theta_m-\theta_n) = -{1\over 2}\sum_{n,m} {\partial_\lambda\big(\langle\partial_\tau\chi_{n,\lambda}|\chi_{m,\lambda}\rangle \langle\partial_\tau\chi_{m,\lambda}|\chi_{n,\lambda}\rangle\big)  \over E_{m,\lambda} - E_{n,\lambda}}(\theta_m-\theta_n) \,,
\end{eqnarray}
Using this,  we find that
\begin{eqnarray}\label{F2fin26}
l_1^{(2)}+ l_{2\,(m\neq n\neq l)}^{(2)} = \sum_{n,m} \Big\{{1\over 2}{\partial_\lambda\big(\langle\partial_\tau\chi_{n,\lambda}|\chi_{m,\lambda}\rangle \langle\partial_\tau\chi_{m,\lambda}|\chi_{n,\lambda}\rangle\big)  \over E_{m,\lambda} - E_{n,\lambda}}\\\nonumber
+3{\langle\partial_\lambda\chi_{m,\lambda}|\chi_{n,\lambda}\rangle\langle \partial_\tau\chi_{n,\lambda}|\chi_{m,\lambda}\rangle\partial_\tau (E_{n,\lambda} - E_{m,\lambda}) \over (E_{n,\lambda} - E_{m,\lambda})^2}    \Big\}(\theta_n-\theta_m).
\end{eqnarray}

Eq. (\ref{F2fin26}) only accounts for $m\neq n\neq l$ terms in Eq. (\ref{F21}) and should be added with $l=m\neq l$, $m=n\neq l$ and $n=l\neq m$ terms. To consider these terms we need to return to the evaluation of the $\omega$ integral in Eq. (\ref{F21}). For $l=m\neq l$ terms we obtain, using Eq. (\ref{hf}),
\begin{eqnarray}\label{F2fin27}
l_{2\,(m=l)}^{(2)} = \sum_{n,m} {\big(\partial_\lambda E_{m,\lambda}\big)\langle\partial_\tau\chi_{m,\lambda}|\chi_{n,\lambda}\rangle \langle\partial_\tau\chi_{n,\lambda}|\chi_{m,\lambda}\rangle (\theta_m-\theta_n) \over (E_{m,\lambda} - E_{n,\lambda})^2}\\\nonumber
= {1\over 2}\sum_{n,m}{\partial_\lambda \big(E_{m,\lambda}-E_{n,\lambda}\big)\langle\partial_\tau\chi_{m,\lambda}|\chi_{n,\lambda}\rangle \langle\partial_\tau\chi_{n,\lambda}|\chi_{m,\lambda}\rangle (\theta_m-\theta_n) \over (E_{m,\lambda} - E_{n,\lambda})^2} \,.
\end{eqnarray}
 Now $l_{2\,(m=l)}^{(2)}$ in Eq. (\ref{F2fin27}) nicely combines with the first term in Eq.(\ref{F2fin26}) to give the full derivative,
\begin{eqnarray}\label{F2fin28}
{1\over 2}\sum_{n\neq m}{\partial\over\partial\lambda}\Big[{\langle\partial_\tau\chi_{m,\lambda}|\chi_{n,\lambda}\rangle \langle\partial_\tau\chi_{n,\lambda}|\chi_{m,\lambda}\rangle \over (E_{m,\lambda} - E_{n,\lambda})}\Big](\theta_m-\theta_n) \,.
\end{eqnarray}

The $m=n\neq l$ and $n=l\neq m$ cases are considered similarly, by reevaluating $\omega$ integrals in Eq. (\ref{F21}). It is easy to see that these two contributions cancel the last term in Eq. (\ref{F2fin26}) and we finally obtain
\begin{equation}\label{F222}
L^{(2)}= {1\over 2}\sum_{n\neq m}\int_0^1 d\lambda {\partial\over\partial\lambda}\Big[{\langle\partial_\tau\chi_{m,\lambda}|\chi_{n,\lambda}\rangle \langle\partial_\tau\chi_{n,\lambda}|\chi_{m,\lambda}\rangle \over (E_{m,\lambda} - E_{n,\lambda})}\Big](\theta_m-\theta_n)    \,.
\end{equation}
Because the integrand is a full derivative over $\lambda$, the value of the integral is the difference of this function at the end points, at $\lambda=1$ and at $\lambda=0$. Using this,  we arrive at Eq. (\ref{S2}).


\begin{thebibliography}{100}
\bibitem{com2} The boundary terms at $\pm\beta/2$, which can potentially appear after integration over $\tau$ by parts, vanish due to periodicity.
\bibitem{com3} The additional, constant term $-mv_s^2/2$ is a pure renormalization of the chemical potential and is irrelevant in the BCS limit for small $v_s$.
\bibitem{new_comm} In Eq. (\ref{vsop}) we have neglected the term $- \sigma_z mv_s^2/ 2$, which renormalizes the chemical potential and is irrelevant for present calculation.
\end{thebibliography}
\end{document}